\documentclass[twocolumn,amsmath,amssymb,prx]{revtex4-2}
\usepackage{latexsym}
\usepackage{graphicx}
\usepackage{psfig}
\usepackage{color}
\usepackage{verbatim}
\usepackage{multirow}
\usepackage[letterpaper, margin=1in]{geometry}
\usepackage{siunitx}
\usepackage[capitalise]{cleveref}
\usepackage{circledsteps}

\begin{document}

\preprint{}
\title{Quasiparticle dynamics in epitaxial Al-InAs planar Josephson junctions}
\author{Bassel Heiba Elfeky$^{1}$}
\thanks{These authors contributed equally.}
\author{William M. Strickland$^{1}$}
\thanks{These authors contributed equally.}
\author{Jaewoo Lee$^{1}$}
\author{James T. Farmer$^{2}$}
\author{Sadman Shanto$^{2}$}
\author{Azarin Zarassi$^{2}$}
\author{Dylan Langone$^{1}$}
\author{Maxim G. Vavilov$^{3}$}
\author{Eli M. Levenson-Falk$^{2}$}
\author{Javad Shabani$^{1}$} \email[]{jshabani@nyu.edu}

\affiliation{$^{1}$Center for Quantum Information Physics, Department of Physics, New York University, New York 10003, USA}
\affiliation{$^{2}$Department of Physics, University of Southern California and Center for Quantum Information Science and Technology, University of Southern California, Los Angeles, California 90089, USA}
\affiliation{$^{3}$Department of Physics and Wisconsin Quantum Institute, University of Wisconsin-Madison, Madison, Wisconsin 53706, USA}
\date{\today}

\begin{abstract}
Quasiparticle (QP) effects play a significant role in the coherence and fidelity of superconducting quantum circuits. The Andreev bound states of high transparency Josephson junctions can act as low-energy traps for QPs, providing a mechanism for studying the dynamics and properties of both the QPs and the junction. Using locally injected and thermal quasiparticles, we study QP loss and QP poisoning in epitaxial Al-InAs Josephson junctions incorporated in a superconducting quantum interference device (SQUID) galvanically shorting a superconducting resonator to ground. We observe changes in the resonance lineshape and frequency shifts consistent with QP trapping into and clearing out of the ABSs of the junctions when the junctions are phase-biased.
By monitoring the QP trapping and clearing mechanisms in time, we find a time-scale of $\mathcal{O}(\SI{1}{\micro s})$ for these QP dynamics, consistent with the presence of phonon-mediated QP-QP interactions. Our measurements suggest that electron-phonon interactions play a significant role in the relaxation mechanisms of our system, while electron-photon interactions and electron-phonon interactions govern the clearing mechanisms. Our results highlight the QP-induced dissipation and complex QP dynamics in superconducting quantum circuits fabricated on superconductor-semiconductor heterostructures.
\end{abstract}

\maketitle

\section{Introduction}
The presence of quasiparticles (QPs) in superconducting materials can prove detrimental to the operation of superconducting quantum circuits, where QP transport and tunneling can cause dissipation \cite{martinis_energy_2009, vepsalainen_impact_2020,martinis_saving_2021, serniak2018, glazman2021, diamond2022} and be a major source of decoherence in charge and parity-based qubits \cite{cheng_topological_2012, rainis2012}. Even at low temperatures, significant densities of non-equilibrium QPs have been observed in superconducting films \cite{aumentado_nonequilibrium_2004, martinis_energy_2009, catelani_quasiparticle_2011, wang_measurement_2014,serniak2018,glazman2021}  due to Cooper pair breaking caused by, for example, leakage of infrared photons \cite{barends_minimizing_2011}, cosmic rays \cite{mcewen_resolving_2022, vepsalainen_impact_2020}, and material defects \cite{kurter_quasiparticle_2022}. At low QP densities, studies have shown that the dominant mechanism for QP relaxation is the diffusive propagation of QPs through the superconductor \cite{riwar_normal-metal_2016}, where they can eventually get trapped in defects or vortices \cite{nsanzineza_trapping_2014, taupin_tunable_2016, patel2017}. At high QP densities, QP recombination involving phonon emission becomes the dominant mechanism for QP relaxation where the emitted phonons can travel through the substrate before getting absorbed by the superconductor and break up Cooper pairs into new QPs \cite{eisenmenger_quantum_1967, otelaja_design_2013, patel2017}.  

Recently, novel superconductor-semiconductor structures have emerged as a promising platform to realize voltage-tunable, wafer-scale superconducting circuit elements such as gatemon qubits \cite{deLange2015, Larsen_PRL, Luthi2018, kringhoj2018, Casparis2018, CasparisBenchmarking, yuan2021, kringhoj_parity2020, Hays2020, hays2021, danilenko2022, hertel2022}, amplifiers \cite{phan2022}, and couplers \cite{casparis2019, Maxim17, sardashti2020, strickland_superconducting_2023}. These material systems have also been studied for their potential application in topological fault-tolerant quantum computing \cite{mayer2019anom, FornieriNature2019, elfeky_local_2021, banerjee_signatures_2022}. In addition to the dissipation and qubit dephasing associated with QPs, in such fault-tolerant schemes, one must conduct braiding operations faster than the QP poisoning time to preserve the parity of the system \cite{cheng_topological_2012, rainis2012}. Thus, understanding the dynamics and effects of QPs in hybrid superconductor-semiconductor structures is vital for the operation of qubits and other superconducting circuit elements on these structures.

\begin{figure*}[ht!]
    \centering
    \includegraphics[width=1.0\textwidth]{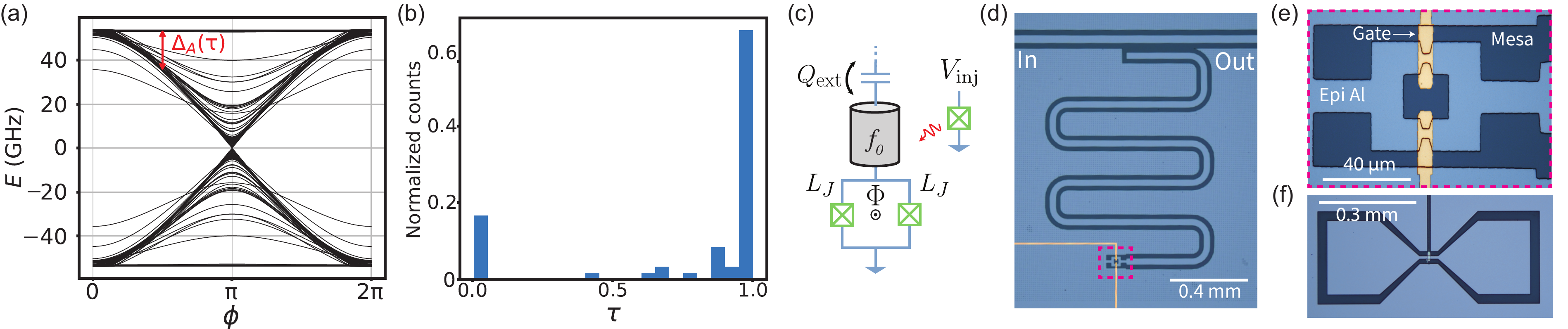}
     \caption{\textbf{Andreev spectrum and device design.} (a) Calculated energy spectrum of the Andreev bound states in a wide Al-InAs junction. The results obtained are for a JJ with width $w = \SI{1}{\micro m}$, normal region length $l = \SI{100}{\nano m}$, length of the superconductor $l_{sc} = \SI{2}{\micro m}$, superconducting gap $\Delta = \SI{220}{\micro eV}$, carrier density $n = 4 \times 10^{11}\SI{}{\centi m^{-2}}$ and effective electron mass $m^* = 0.04m_{e}$ where $m_{e}$ is the electron mass. 
     (b) A histogram of the transparencies of the ABS modes in (a) extracted from fitting the modes to \cref{eq:gen_abs}. 
     (c) Circuit diagram of the CPW-SQUID device. A transmission line resonator with frequency $f_{0}$ is coupled capacitively to a feedline characterized by an external quality factor $Q_{\mathrm{ext}}$ and directly connected to a superconducting loop with two Josephson junctions, each with inductance $L_{J}$. A flux $\Phi$ threads the loop. Separately, a two-terminal ``injector'' Josephson junction is placed \SI{1.6}{\milli m} away from the SQUID, serving to inject QPs to the circuit by biasing one terminal with voltage $V_{\mathrm{inj}}$ and grounding the other terminal. (d) Optical images of the CPW-SQUID device, (e) the SQUID, and (f) the injector junction where the etched mesa is shown in dark blue, epitaxial Al in light blue, and gates in gold.
     }
    \label{fig:fig1}
\end{figure*}

In semiconductor-based Josephson junctions (JJs), supercurrent is carried by electrons and holes in conduction channels mediated by Andreev reflection \cite{beenaker1992}. Coherent Andreev reflections result in sub-gap Andreev bound states (ABSs), with each channel forming a pair of ABSs, having energy given by:

\begin{equation}
E_{A}^{\pm}(\phi) = \pm\Delta \sqrt{1-\tau\sin^2(\phi/2)}
\label{eq:gen_abs}
\end{equation} 

where $\phi$, $\tau$, and $\Delta$ are the phase difference across the junction, transparency, and superconducting gap, respectively. The energy spectrum of the ABSs in an Al-InAs JJ with width $w =\SI{1}{\micro m}$ is shown in \cref{fig:fig1}(a) obtained using tight-binding simulations with realistic parameters (see Appendix A). The simulations present an Andreev spectrum with a large number of modes and a few long junction modes, diffusive modes detached from the continuum at $\phi = 0$, present due to the wide nature of the planar JJ \cite{dartiailh_missing_2021, elfeky_reemergence_2022}. Fitting the modes in the Andreev spectrum to \cref{eq:gen_abs}, we extract the transparencies ($\tau$) of these modes and plot them as a histogram in \cref{fig:fig1}(b) where we observe the majority of modes having high transparency and a small fraction of modes having intermediate transparency. Typically, at low temperatures, ABSs with energy $E_A^-$ are occupied, and those with energy $E_A^+$ are unoccupied. QPs, which have energy $\gtrsim\Delta$, can relax from the quasicontinuum into one of the Andreev levels, which effectively has a trap of depth $\Delta_{A}(\phi,\tau) = \Delta- E^{+}_{A}(\phi,\tau)$, as denoted in \cref{fig:fig1}(a). Since ABSs with $E^{+}_A$ carry current in the opposite direction to those ABSs with $E^{-}_A$, a QP falling into the positive energy ABS poisons the conduction channel resulting in the channel carrying zero supercurrent, a process known as QP poisoning. The resulting decrease in supercurrent, and the corresponding increase in inductance, can be readout using standard dispersive measurement techniques as demonstrated in Al-nanobridge JJs \cite{levensonfalk2014, farmer2021, farmer2022} and also in Al-InAs Josephson nanowires where the spin state of a QP trapped in the Andreev level is used as the basis state of an Andreev spin qubit \cite{Hays2020,hays2021}. Trapped QPs can then be excited or cleared out of their Andreev traps by applying a high frequency $f_\mathrm{clear}$ clearing tone with $h f_\mathrm{clear}>\Delta_{A}$. 

In this work, we study the trapping and clearing of QPs in epitaxial Al-InAs planar Josephson junctions embedded in a superconducting quantum interference device (SQUID) that electrically shorts a coplanar waveguide (CPW) resonator to ground. We show that by increasing the QP density through local QP injection or by raising the temperature, QPs can trap in the ABSs of the junctions at non-zero phase bias and examine their effect on the resonance lineshape. Further, we study the QP trapping and clearing dynamics by pulsing a clearing tone and measuring the time scales associated with the relaxation and excitation of QPs into and from the traps.

\section{Device design and measurement setup}

The devices used in this study are fabricated on a superconductor-semiconductor (Al-InAs) heterostructure grown via molecular beam epitaxy \cite{Shabani2016, Kaushini2018, Yuan2020, strickland2022}. The weak link of the JJ is an InAs 2DEG grown near the surface and contacted \textit{in-situ} by a thin Al film. The heterostructure is grown on a \SI{500}{\micro m} thick InP substrate. We use an III-V wet etch to define the microwave circuit, and an Al wet etch to define the JJ. A blanket layer of AlO$_\text{x}$ is then deposited as a gate dielectric followed by a patterned layer of Al for gate electrodes which are all kept grounded during the measurements. Further details on the growth and fabrication are provided in Appendix B.

Our device consists of a hanger $\lambda/4$ CPW resonator (with geometric inductance $L_{0} = \SI{1.573}{\nano H}$ and capacitance $C_{0} = \SI{414}{\femto F}$) coupled capacitively to a common feedline, characterized by an external quality factor $Q_\mathrm{ext} \approx$ 1440. The resonator is shorted to ground through a SQUID with two symmetric JJs that are $w \approx\SI{4}{\micro m}$ wide and $l \approx\SI{100}{\nano m}$ long. Coupled to the same feedline is a bare test resonator used for characterizing the Al film and two other CPW-SQUID devices, which we do not discuss in this work. Around the microwave circuit, we etch $2\times\SI{2}{\micro m}$ holes in the ground plane on a $\SI{10}{\micro m}$ grid to act as flux-pinning holes. Near the microwave circuit, \SI{1.6}{\milli m} away from the SQUID, we have a two-terminal JJ, fabricated on the same chip with the same junction structure, used for injecting QPs. By dc voltage-biasing the ``injector'' JJ above twice the superconducting gap, $e V_\mathrm{bias}>2\Delta$, QPs are generated near the injector JJ area. These QPs can then relax to the gap edge or recombine, emitting phonons that propagate through the substrate breaking Cooper pairs successively, increasing the density of QPs in the circuit. A similar QP injection mechanism has been used in Refs. \citenum{patel2017, iaia_phonon_2022, bagerbos2022}. A circuit diagram of the CPW-SQUID device and injector junction is shown in \cref{fig:fig1}(c) with optical micrographs shown in \cref{fig:fig1}(d)-(f).

The chip is measured in a dilution fridge at a temperature $T = \SI{15}{\milli K}$, mounted on the mixing chamber in a QDevil QCage, a microwave cavity sample holder with EMC-tight superconducting shielding. An out-of-plane magnetic field is applied to the chip using a superconducting coil placed inside the QCage shielding. All DC lines go through a QDevil Qfilter, a low-pass filter with a resistance of $\SI{1.7}{\kilo \Omega}$, such that the reported applied voltage bias to the injector $V_\mathrm{inj}$ is applied across the filter and the injector junction in series. We expand on the measurement setup in Appendix C. Unless specified otherwise, the measurements are all performed at a photon number of $\approx 50$. We first measure the complex transmission $S_{21}$ and fit the resonant response in transmission to extract the resonant frequency $f_{r}$ and the internal quality factor $Q_\mathrm{int}$ using a fitting procedure outlined in Ref. \citenum{Probst}. Time-domain measurements are performed by applying a probe tone $f$ = $f_{r}(\Phi/\Phi_{0})$ and IQ demodulating, \SI{22}{\mega Hz} low pass-filtering and digitizing at \SI{500}{\mega Sa/s} the outgoing signal.

\begin{figure}[ht!]
    \centering
    \includegraphics[width=0.45\textwidth]{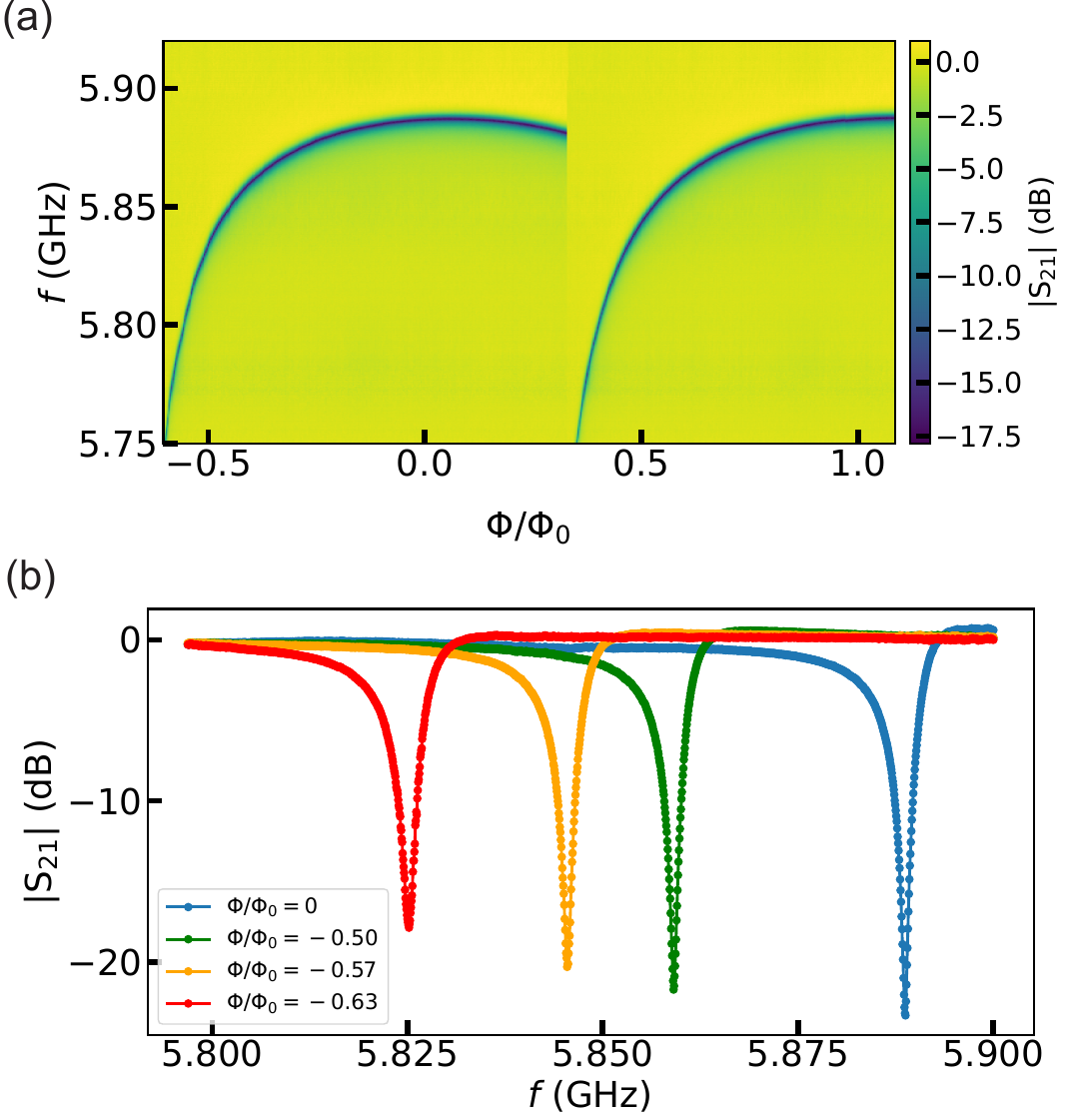}
         \caption{\textbf{Flux tuning.} 
         (a) Magnitude of complex transmission $|S_{21}|$ as a function of magnetic flux $\Phi/\Phi_{0}$ with linecuts shown in (b).
         }
    \label{fig:fig2}
\end{figure}

\section{SQUID flux biasing}

Applying an external flux $\Phi$ to the SQUID, we can phase bias the JJs such that the junctions' phase bias is $\phi = \pi\Phi/\Phi_{0}$ where $\Phi_{0}$ is the magnetic flux quantum. At half-flux ($\Phi/\Phi_{0}=0.5$) the phase of the two junctions is $\phi = \pi/2$ which, following \cref{eq:gen_abs}, creates a maximum Andreev trap of approximately $\Delta_{A} (\tau = 1)/h \approx \SI{15}{\giga Hz}$ for a superconducting gap $\Delta = \SI{210}{\micro eV}$.
In \cref{fig:fig2}(a), we show a colormap of the response in $|S_{21}|$ as a function of $\Phi/\Phi_{0}$ with linecuts of the resonance shown in \cref{fig:fig2}(b). The SQUID response exhibits hysteretic modulation with $f_{r}$ continuing to decrease past $\Phi/\Phi_{0} = 0.5$, which can be attributed to the non-sinusoidal CPR of the highly transparent Al-InAs junctions \cite{levenson-falk_nonlinear_2011,mayer2019anom, Fabrizio17}. We provide further details on the SQUID oscillation periodicity and flux calibration in Appendix D. The CPW-SQUID has a resonant frequency of $f_{r} (\Phi/\Phi_{0}=0.0) = \SI{5.89}{\giga Hz}$ and $f_{r} (\Phi/\Phi_{0}=0.5) = \SI{5.86}{\giga Hz}$ corresponding to Josephson inductance of $L_{J}(\Phi/\Phi_{0}=0.5) = \SI{0.190}{\nano H}$ and a critical current of $I_{c}(\Phi/\Phi_{0}=0.5) = \SI{1.73}{\micro A}$ (see Appendix E) similar to values reported using dc measurements on Al-InAs heterostructures \cite{nichele_relating_2020, dartiailh_phase2021}.

\begin{figure}[ht!]
    \centering
    \includegraphics[width=0.45\textwidth]{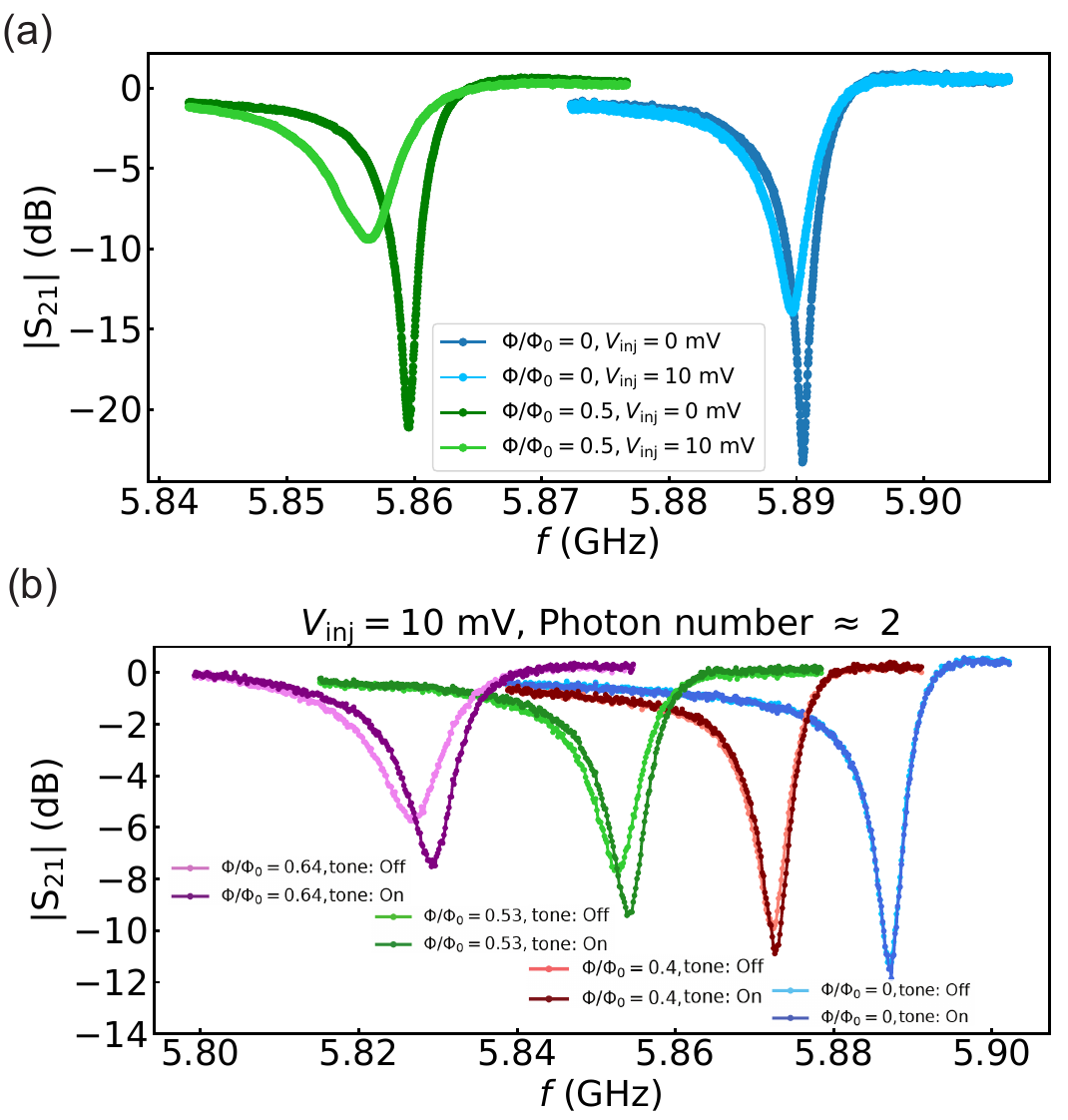}
         \caption{\textbf{Effect of quasiparticle injection and clearing.} 
         (a) Linecuts for zero and half-flux at zero and finite injector bias, $V_\mathrm{inj}$. The half-flux case shows a more pronounced broadening and frequency shift upon QP injection.
         (b) Linecuts at different flux values at $V_\mathrm{inj} = \SI{10}{\milli V}$ with and without an applied clearing tone of frequency $f_\mathrm{clear} = \SI{18}{\giga Hz}$. The finite flux cases shows some of its lineshape restored with the application of the clearing tone, while the zero-flux case shows no change.
         }
    \label{fig:fig3}
\end{figure}

Comparing the linecuts in \cref{fig:fig2}(b), we see that the resonance gets slightly shallower and broader as the applied flux increases. This effect is commonly seen in flux-tuned SQUIDs and can be attributed to a finite subgap resistance in the junctions or to thermal noise \cite{palacios-laloy_tunable_2008, sandberg2008}. We note that quasiparticle trapping in the ABSs result in a similar response to an applied flux. We apply a sufficiently high-frequency clearing tone $f_\mathrm{clear} = \SI{18}{\giga Hz}$ on the feedline input to excite any trapped QPs out of ABS traps. In the presence of QP trapping, we expect the clearing tone to cause the resonance shape to become deeper and narrower as trapped QPs are cleared from the junctions. The frequency $f_\mathrm{clear} = \SI{18}{\giga Hz}$ is chosen since it is near the third harmonic of the $\lambda/4$ resonator where the admittance of the CPW is peaked. Upon the application of the clearing tone, however, we observe no noticeable change to the resonance shape, indicating the absence of QP trapping at base temperature with no deliberate QP injection. This lack of noticeable QP poisoning effects is possibly due to a variety of reasons which include a low non-equilibrium QP density near the SQUID and/or a lack of sensitivity to QP poisoning events given our resonance linewidth; a detailed discussion about this is provided in Appendix E. 

We increase the QP density in the circuit by applying a finite voltage bias to the injector junction $V_\mathrm{inj}$. As seen in \cref{fig:fig3}(a), at zero- and half-flux with $V_\mathrm{inj} = \SI{10}{\milli V}$ the resonance lineshape becomes shallower and broader and exhibits a shift towards negative frequency, where the effect is considerably more pronounced in the half flux case. Two main factors can contribute to this effect: an increase in dissipation due to bulk QP transport in the bulk superconducting film and/or an increase in QP trapping in the ABSs of the junctions in the SQUID. In \cref{fig:fig3}(b), we plot the resonance curves at $V_\mathrm{inj} = \SI{10}{\milli V}$ with and without an applied clearing tone at different flux-biases. The measurements in \cref{fig:fig3}(b) are performed at a photon number of $\approx 2$ to avoid any clearing effects from the readout tone $f$. We observe that at zero-flux, the clearing tone does not affect the resonance shape, while at finite flux, the application of the clearing tone results in the resonance becoming sharper and deeper, exhibiting a positive frequency shift, consistent with the clearing of QPs from the ABS traps. These results indicate the presence of QP trapping in ABSs with QP injection when the junctions are phase-biased.

\section{QP induced dephasing and dissipation}

\begin{figure*}[htp]
    \centering
    \includegraphics[width=0.95\textwidth]{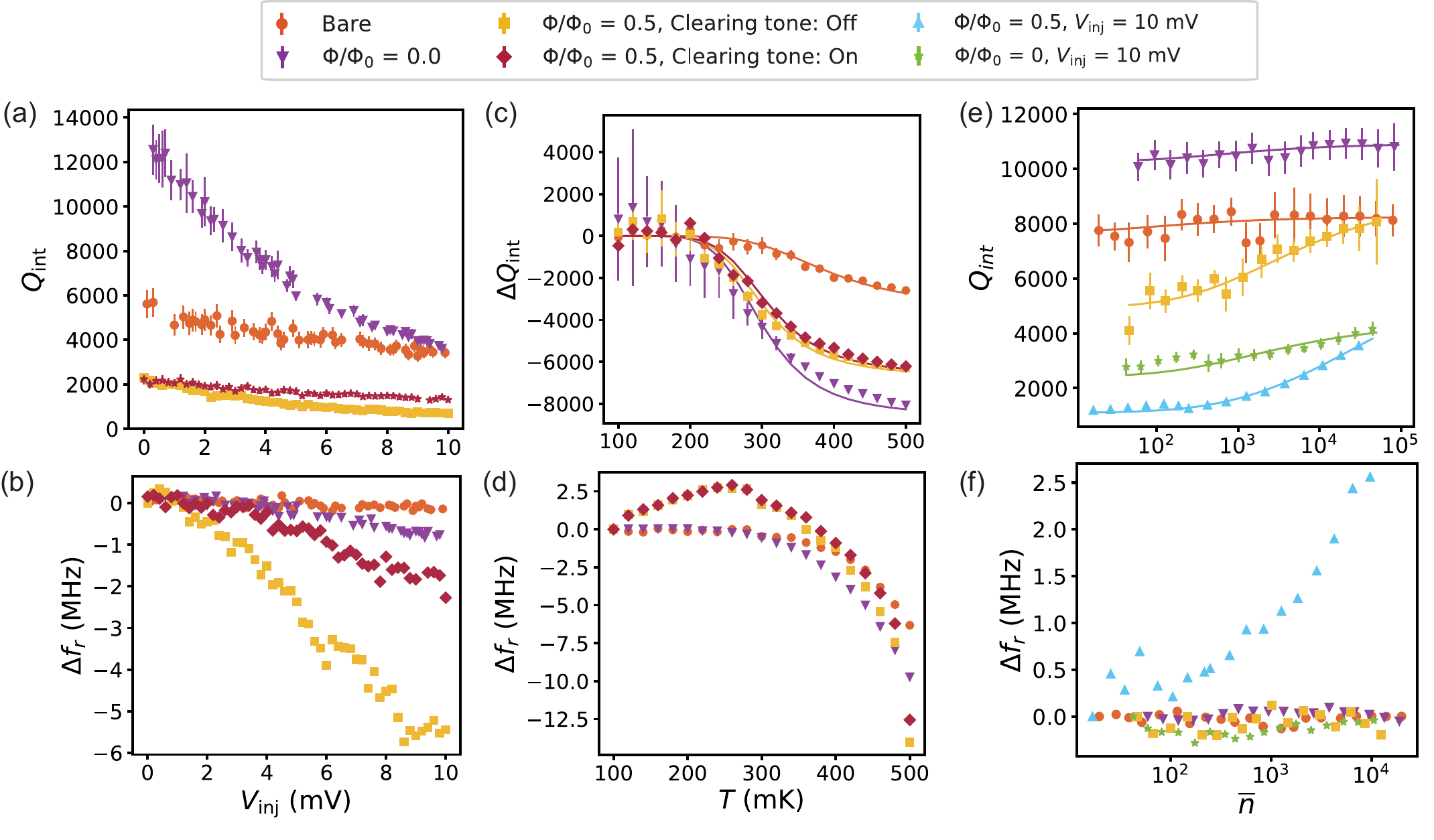}
         \caption{\textbf{Microwave loss due to quasiparticle effects.} Internal quality factor $Q_\mathrm{int}$ and frequency shift $\Delta f_{r}$ as a function of: (a)-(b) voltage bias of the injector junction $V_\mathrm{inj}$ at $T = \SI{15}{\milli K}$, (c)-(d) temperature $T$ without QP injection ($V_\mathrm{inj} = \SI{0}{\milli V}$) and (e)-(f) photon number $\bar n$ at $T = \SI{15}{\milli K}$. The $\Delta f_{r}$ is calculated by subtracting each data set's leftmost $f_{r}$ value on the horizontal axis. Solid lines show fits to models for the dependence of $Q_\mathrm{int}$ on $T$ and $\bar{n}$.}
    \label{fig:fig4}
\end{figure*}

By varying the quasiparticle density near the junctions, we examine the effects of QPs on the CPW-SQUID device in terms of the internal quality factor $Q_\mathrm{int}$ and frequency shift $\Delta f_{r}$. We increase the QP density by applying a voltage bias across the injector junction $V_\mathrm{inj}$ or by raising the temperature $T$, and also examine the clearing of QPs by the microwave photons of the readout tone. We include measurements of a bare resonator on the same chip as a reference to differentiate between effects resulting from QP trapping in the ABSs of the junctions and from dissipation due to QP transport in the superconducting film. We note that the bare resonator is located \SI{1.6}{\milli m} further from the injector junction than the CPW-SQUID device; thus, we expect that at a given $V_\mathrm{inj}$ there is a lower density of injected QPs near the bare resonator compared to near the CPW-SQUID device \cite{patel2017}. It is worth mentioning that in the presence of QP trapping in ABSs, changes in $Q_\mathrm{int}$ are associated more with dephasing of the resonant frequency and the associated frequency noise rather than an increase in dissipative loss \cite{levenson-falk_nonlinear_2011}. 

We again vary $V_\mathrm{inj}$ as shown in \cref{fig:fig4}(a) and find that $Q_\mathrm{int}$ of the bare resonator decreases due to dissipation from the increased density of QPs in the superconducting Al film. The $Q_\mathrm{int}$ response of the CPW-SQUID at $\Phi/\Phi_{0} =0.0$ and $\Phi/\Phi_{0} =0.5$, shows a pronounced dependence on $V_\mathrm{inj}$. While QP-induced dissipation in the resonator of the CPW-SQUID is expected to be enhanced due to its position relative to the injector junction, the presence of the junctions can introduce processes such as QP transport via resistive conduction channels in the junction and QP trapping (for the $\Phi/\Phi_{0} =0.5$ case) which can contribute to the pronounced response in $Q_\mathrm{int}$. Examining the $V_\mathrm{inj}$ dependence on the shift in resonant frequency $\Delta f_{r}$ presented in \cref{fig:fig4}(b), we see that the $\Phi/\Phi_{0} = 0.5$ case shows a significantly more drastic trend than $\Phi/\Phi_{0} =0.0$. When a high-frequency clearing tone is applied to the $\Phi/\Phi_{0} =0.5$ case, we find $\Delta f_{r}$ shows a significant change, as the shift decreases from \SI{5.7}{\mega Hz} to \SI{2.1}{\mega Hz} at $V_\mathrm{inj}$, along with $Q_\mathrm{int}$ increasing by a factor of $\approx 2$. This maximum $\Delta f_{r}$, with the application of the clearing tone, not being closer to the $\Phi/\Phi_{0} =0.0$ case, could indicate inefficiency in the clearing process, as we discuss further in Appendix F. We note that applying a clearing tone to the CPW-SQUID at $\Phi/\Phi_{0} = 0.0$ does not effect $f_{r}$ or $Q_\mathrm{int}$.

An increase in the equilibrium QP density can be achieved by increasing the temperature. As we vary the temperature, we observe that $Q_\mathrm{int}$ is roughly unchanged until $T\approx \SI{200}{\milli K}$ for the CPW-SQUID and $T\approx \SI{300}{\milli K}$ for the bare resonator, at which point $Q_\mathrm{int}$ starts to decrease as shown in \cref{fig:fig4}(c). The dependence is again seen to be more pronounced in the CPW-SQUID than the bare resonator. To analyze the trend in $Q_\mathrm{int}$, it is useful to write the total $Q_\mathrm{int}$ as an inverse sum of terms which capture the temperature $T$ and photon number $\bar n$ dependencies separately, being 

\begin{equation}
    \frac{1}{Q_\mathrm{int}(T, \bar n)} = \frac{1}{Q_0}+\frac{1}{Q(T)} +\frac{1}{Q(\bar n)}
    ,
    \label{eqn:q}
\end{equation}
where $Q_0$ describes losses independent of $T$ and $\bar{n}$. These terms can be interpreted as contributions to loss due to thermal QPs, in the case of $Q(T)$, and QP excitation due to the readout tone, $Q(\bar n)$.
We consider a model for the temperature-dependent surface impedance following the Mattis-Bardeen theory, where the temperature-dependent internal quality factor is given by
\begin{equation}
    Q(T) = Q_{\mathrm{QP}, 0} \frac{e^{\Delta/k_B T}}{\sinh\left(\frac{hf_r}{2k_BT} \right)K_0\left(\frac{hf_r}{2k_BT}\right)},
    \label{eqn:qqp}
\end{equation}
 as described in Refs. \citenum{Gao2008, zmuidzinas_superconducting_2012}, where $Q_{\mathrm{QP},0}$ is the inverse linear absorption by quasiparticles, $\Delta = \SI{210}{\micro eV}$ is the superconducting gap of aluminum, and $h$ is Planck's constant. The resonant frequency $f_r$ is set its value at $T = \SI{100}{\milli K}$. The only free parameters in the fit are $Q_{\mathrm{QP},0}$ and the inverse sum $1/Q_0+1/Q(\bar n)$. The fit to \cref{eqn:qqp} for each data set is shown in \cref{fig:fig4}(c). Here, we plot the results in terms of the change in internal quality factor $\Delta Q_\mathrm{int}$ to accentuate the temperature dependence. For the bare resonator, we find that the inverse linear absorption for the bare resonator is $Q_{\mathrm{QP},0} = 3.74$. This benchmarks quasiparticle dissipation due to the superconducting Al film in the resonator. We find that in the SQUID-CPW device, at $\Phi/\Phi_0 = 0.0$, $Q_\mathrm{QP, 0}$ decreases to 1.28, lower than that of the bare resonator by a factor of 66\%. This implies that other temperature-dependent factors play a role with the addition of the SQUID loop to the resonator, such as the critical current carried by the junctions, the number of resistive channels, and the induced superconducting gap of the junctions. Further, we find that as the SQUID is flux-biased to $\Phi/\Phi_0 = 0.5$, $Q_\mathrm{QP, 0}$ further decreases to 1.02. This decrease is expected and corresponds to the effect of QP trapping. An interesting observation is the recovery of $Q_\mathrm{QP, 0}$ upon the application of a clearing tone, where $Q_\mathrm{QP, 0}$ = 1.27, which is almost the value of $Q_\mathrm{QP, 0}$ for $\Phi/\Phi_0 = 0.0$. Shifting our focus to the $\Delta f_{r}$ dependence, presented in \cref{fig:fig4}(d), the $\Phi/\Phi_0 = 0.5$ case shows a gradual increase in $\Delta f_{r}$ up until $T\approx$ \SI{250}{\milli K}, rather than staying constant like the bare resonator and the CPW-SQUID at $\Phi/\Phi_0 = 0.0$.
This increase in frequency with temperature is consistent with the suppression of QP trapping due to the growing population of thermal phonons with hotter QPs being less likely to relax and get trapped in lower-energy trap states or an increase in QP clearing mediated by phonon absorption \cite{levensonfalk2014, farmer2022}.
While the equilibrium phonons do not provide enough energy ($\approx \SI{5.2}{\giga Hz}$ at $T = \SI{250}{\milli K}$) for the QPs to be cleared from the deepest Andreev traps directly to the continuum, the presence of intermediate transparency modes as seen in \cref{fig:fig1} and mode-to-mode coupling \cite{elfeky_reemergence_2022}, can also support excitations of QPs from deep high-transparency modes to shallower modes with less transparency and eventually to the continuum through multiple transitions.
At temperatures above \SI{250}{\milli K}, the effect of the rising QP density dominates, and $\Delta f_{r}$ begins to decrease in a similar fashion to the bare resonator and SQUID with $\Phi/\Phi_0 = 0.0$. 

We next consider the dependence of $Q_\mathrm{int}$ and $\Delta f_{r}$ on the photon number $\bar n$ of the readout tone, presented in \cref{fig:fig4}(e) and (f). The bare resonator shows a $Q_\mathrm{int}$ that is approximately independent of $\bar n$, suggesting that the loss is not limited by two-level systems (TLSs) \cite{wenner_surface_2011, sage_study_2011, zmuidzinas_superconducting_2012}. At $V_\mathrm{inj}$ = \SI{0}{\milli V}, a similar $\bar n$ independent loss is observed in the CPW-SQUID at $\Phi/\Phi_{0} =0.0$ but not at $\Phi/\Phi_{0} =0.5$. The $\bar n$ dependence observed is more pronounced at $V_\mathrm{inj} = \SI{10}{\milli V}$ for $\Phi/\Phi_0 = 0.5$. This $\bar n$ dependent pattern is also seen in $\Delta f_{r}$ for the $\Phi/\Phi_0 = 0.5$ case at $V_\mathrm{inj} = \SI{10}{\milli V}$ where $\Delta f_{r}$ decreases by \SI{2.5}{\mega Hz} from high to low $\bar n$. These results are consistent with the clearing of trapped QPs mediated by the photons of the readout tone. Similar to clearing mediated by thermal phonons,
the absorption of microwave photons, even photons that carry less energy than the Andreev depth ($f = $ 5.86--$\SI{5.89}{\giga Hz}$ compared to $\Delta_{A} (\tau = 1)/h \approx \SI{15}{\giga Hz}$), can contribute to the clearing of QPs through multi-photon transitions and excitations to subsequently higher energy Andreev levels at high enough $\bar n$ as evident from \cref{fig:fig4}(f). On the other hand, at low $\bar n$, the readout tone does not provide enough power to support quasiparticle clearing from the ABSs, leading to a saturation in the low power frequency shift and internal quality factor. This suggests that, at low temperatures, electron-photon interactions can be a contributing mechanism for QP clearing, consistent with results reported for Al-nanobridge junctions \cite{farmer2022}.

We consider the changes in $Q_\mathrm{int}$ as a function of photon number to be associated with the changing densities of trapped and mobile QPs via excitation from the readout tone. Considering a model detailed in Ref. \citenum{grunhaupt2018}, one can write down a kinetic equation that includes the rates of QP recombination $\Gamma_{ml}$ (where $m$ and $l$ denote mobile and localized QPs respectively), localization $\Gamma_\mathrm{loc}$, excitation $\Gamma_\mathrm{ex}$, and generation $g$. We fit the photon number dependence of the internal quality factor in terms of the excitation of trapped QPs using \cref{eqn:q} and the following expression for $Q(\bar n)$

\begin{equation}
    Q(\bar{n}) = \frac{1}{\beta} \left[1+\frac{\gamma\bar n}{1+ \frac{1}{2}(\sqrt{1+4\gamma \bar n}-1)} \right],
    \label{eqn:power}
\end{equation}
where $\beta$ is a prefactor determining the dependence on photon number and is proportional to the ratio of QP localization and recombination (specifically of mobile and localized QPs) $\Gamma_{loc}/\Gamma_{ml}$. The factor $\gamma$ is defined as $\gamma = 2\Gamma_\mathrm{loc}\Gamma_0 / g_m$ where $\Gamma_0$ describes the strength of the coupling between an excitation tone and the QPs, related to $\Gamma_\mathrm{ex}$ through $\Gamma_\mathrm{ex} = \Gamma_0 \bar{n}$, and $g_m$ is the generation rate of mobile QPs. Fitting the data for $Q_\mathrm{int}$ to \cref{eqn:power}, we find that the bare resonator has $\beta = \SI{9.23e-6}{}$, while in the CPW-SQUID device, $\beta = \SI{5.95e-6}{}$ at $\Phi/\Phi_0 = 0.0$. We find that $\beta$ increases to $\SI{9.01e-5}{}$ at $\Phi/\Phi_0 = 0.5$. If we assume the rate for QP recombination is independent of applied flux, this corresponds to the increase in the rate of QP localization at $\Phi/\Phi_0 = 0.5$ of a factor of 15, which we expect to be due to the deepening of the ABS traps with applied flux. With QP injection at $V_\mathrm{inj} = \SI{10}{\milli V}$, we find that $\beta$ increases to $\beta = \SI{1.86e-4}{}$ at $\Phi/\Phi_0 = 0.0$ and to $\beta = \SI{7.47e-4}{}$ at $\Phi/\Phi_0 = 0.5$ compared to without QP injection. These results imply that QP localization rate increases with an applied finite $V_\mathrm{inj}$ increasing the background quasiparticle density. We also anticipate that by increasing the number of mobile and trapped QPs, the rate of recombination should increase as well, possibly explaining why the increase in $\beta$ for the SQUID with $\Phi/\Phi_0 = 0.5$ is not as drastic as that for the $\Phi/\Phi_0 = 0.0$ case. In terms of fitting for $\gamma$, we find that several different values for $\gamma$ yield similar fit results. The value of $\gamma$ used to obtain the fits for the bare resonator is 0.013, that for the SQUID with $\Phi/\Phi_0 = 0.0$ is 0.002, and that for the SQUID with $\Phi/\Phi_0 = 0.5$ is 0.001.

\section{Trapping and Clearing Dynamics}

\begin{figure*}[htpb!]
    \centering
    \includegraphics[width=0.95\textwidth]{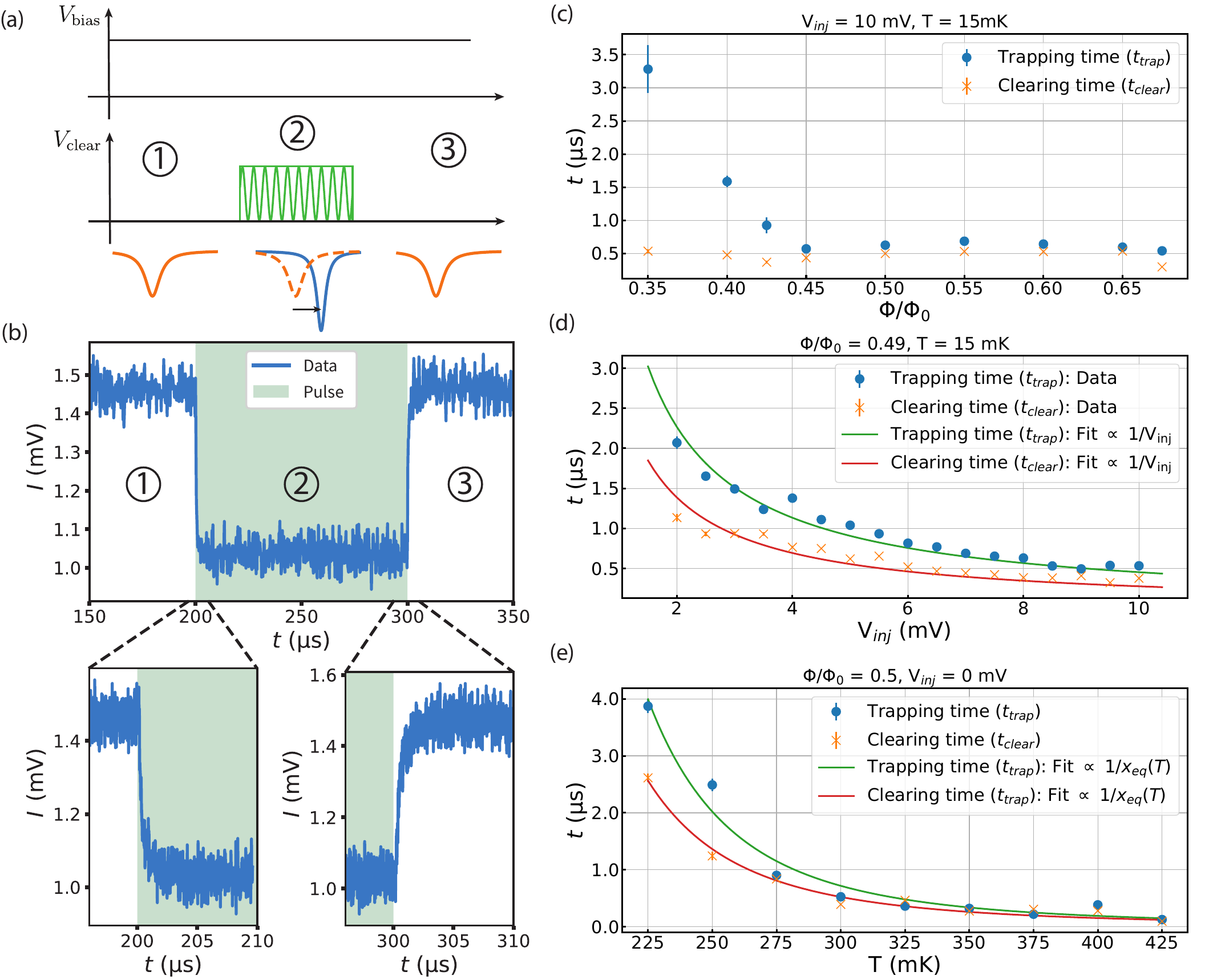}
    \caption{\textbf{Quasiparticle trapping and clearing.} (a) Protocol quasiparticle clearing: \Circled{1} While the in-phase (I) and quadrature (Q) components are measured, a constant $V_\mathrm{inj} = \SI{10}{\milli V}$ being applied such that a finite QP density populates the traps states . \Circled{2} A \SI{100}{\micro s} long clearing tone pulse with $f_\mathrm{clear} = \SI{18}{\giga Hz}$ is applied which clears out the QPs. \Circled{3} The clearing tone pulse ends and QPs fall back into the trap states. (b) Example measurement of the I response over time $t$. The region where the pulse is maximum is shaded in green. The clearing time $t_\mathrm{clear}$ and the trapping $t_\mathrm{trap}$ are extracted from an exponential fit of the I or Q response to the start (bottom left plot) and end of the pulse (bottom right plot), respectively. Trapping and clearing times as a function of $\Phi/\Phi_0$, $V_\mathrm{inj}$ and temperature $T$ are shown in (c), (d), and (e), respectively.}
    \label{fig:fig5}
\end{figure*}

In order to further probe the dynamics of QP trapping and clearing, we monitor the response of the resonance in time while pulsing a clearing tone. The protocol used is shown in \cref{fig:fig5}(a): a constant $V_\mathrm{inj}$ is applied, constantly injecting QPs into the system and resulting in a finite population of QPs occupying the trap states. An applied high-frequency clearing tone $f_\mathrm{clear} = \SI{18}{\giga Hz}$ is mixed with a \SI{100}{\micro s} wide square pulse output by an arbitrary waveform generator with a sampling rate of \SI{1}{\giga Sa/s}. The output signal's two quadratures, $I$ and $Q$, are measured as the clearing tone is pulsed and is averaged over 10,000 runs. \cref{fig:fig5} (b) shows an example of the response of $I$ measured in time where a decay is observed when the pulse begins ($t = \SI{200}{\micro s}$), corresponding to a shift in the resonant frequency upon the application of the clearing tone and the clearing of QPs. The resonance reaches a steady state with the clearing tone applied, and when the clearing tone pulse ends ($t = \SI{300}{\micro s}$), $I$ decays to its initial values as QPs start reoccupying the traps. We fit the quadrature responses at the start and end of the pulse in time to an exponential decay and extract two time constants: the clearing time and the trapping time. The clearing time $t_\mathrm{clear}$ is the time constant associated with the excitation of QPs from the Andreev traps mediated by the applied clearing tone, while the trapping time $t_\mathrm{trap}$ is the time constant associated with the relaxation of QPs from the continuum into the Andreev traps once the pulse ends. If the clearing and trapping events are uncorrelated, the clearing and trapping times correspond to the rate for individual QPs to clear or trap.

We first tune the trap depth $\Delta_{A}$ via flux $\Phi/\Phi_{0}$ and monitor the effect on $t_\mathrm{trap}$ and $t_\mathrm{clear}$ at $V_\mathrm{inj} = \SI{10}{\milli V}$ as seen in \cref{fig:fig5}(c). Between $\Phi/\Phi_{0} = 0.35$ and $0.675$, we find trapping times of order $ \SI{1}{\micro s}$ which decrease as a function of $\Phi/\Phi_0$. The trend implies that deeper Andreev traps are more likely to trap QPs, consistent with the electron-phonon relaxation mechanism for QP trapping \cite{levensonfalk2014, patel2017, farmer2022}.

On the other hand, the clearing time is expected to be less dependent on flux and should depend on the clearing tone power \cite{levensonfalk2014}. This is because the clearing tone should provide sufficient energy to excite QPs out of the deepest Andreev traps, if not directly to the continuum, then to lower transparency modes carrying significantly less supercurrent and correspondingly contribute less to the response. This is consistent with the data in \cref{fig:fig5}(c) where $t_\mathrm{clear}$ remains around $\sim \SI{500}{\nano s}$ at all fluxes.  We note that below $\Phi/\Phi_{0} = 0.35$, the signal-to-noise ratio is too low to obtain an accurate fit. The fact that the clearing and trapping times are of similar magnitude might also contribute to the inefficiency of the clearing tone as discussed for the $\Delta f_{r}$ results shown in \cref{fig:fig4}(d), where Andreev states can remain partially occupied in the presence of the clearing tone as the system reaches a steady state where trapping and clearing events balance, providing a finite QP occupation probability even in the presence of the clearing tone.

Assuming QP trapping and clearing are independent processes and the clearing tone can be treated as a large bath of photons, such systems are typically described by a Markovian model \cite{farmer2022}. In this case, the trapping and clearing rates of individual QPs are expected to be independent of the QP density. In \cref{fig:fig5}(d), we present the dependence of $t_\mathrm{trap}$ and $t_\mathrm{clear}$ on $V_\mathrm{inj}$ at $\Phi/\Phi_{0} = 0.49$. Without QP injection and up to $V_\mathrm{inj} = \SI{2}{\milli V}$, we do not observe a noticeable trapping/clearing response in I or Q (see Appendix G). Above $V_\mathrm{inj} = \SI{2}{\milli V}$, we find that as the QP density increases, $t_\mathrm{trap}$ decreases by a factor of 4 at $V_\mathrm{inj} = \SI{10}{\milli V}$. This corresponds to the system recovering to a steady state faster, suggesting faster trapping rates when the QP density is higher. The QP density-dependent trapping rates can be attributed to a rise in phonon populations at higher QP densities. An increase in QP recombination at higher QP densities results in the emission of high-energy ($\geq 2\Delta_{0}$) phonons which can then enhance the number of stimulated emission events for QPs and consequently help increase the QP relaxation and trapping rates, decreasing the trapping times. Further, QP relaxation mediated by phonon emission, typically low-energy phonons, can also contribute to the increasing phonon population and, consequently to the increase in trapping rates. Interestingly, $t_\mathrm{clear}$ also exhibits a dependence on $V_\mathrm{inj}$ decreasing by a factor of 3 in the measured $V_\mathrm{inj}$ range. This $t_\mathrm{clear}$ dependence implies that QPs are more likely to be cleared out of the Andreev traps at higher QP densities. This is possibly also due to the rise in phonon populations, generated through QP recombination and phonon-mediated QP relaxation, that can then participate in the excitation of QPs out of the deepest Andreev traps either directly to the continuum or through multiple transitions, increasing the clearing rate which corresponds to a decrease in the clearing times. These results suggest that electron-phonon interactions play a considerable role in QP dynamics even at low temperatures. Further, we find that both $t_\mathrm{trap}$ and $t_\mathrm{clear}$ fit inversely proportionally to $V_\mathrm{inj}$ as shown with the fits  in \cref{fig:fig5}(d). The origin of this dependence can be understood by considering a phenomenological model for the QP density as discussed in Appendix H.

Finally, we examine the effect of raising the temperature $T$ without QP injection ($V_\mathrm{inj} = \SI{0}{\milli V}$), corresponding to increasing the thermal equilibrium QP density $x_\mathrm{eq}$ which varies as
\begin{equation}
x_\mathrm{eq}(T) = \sqrt{\frac{2 \pi k_B T}{\Delta} } \exp\left(\frac{-\Delta}{k_B T}\right)  
\label{eq:eq_thermal}
\end{equation} where $k_{B}$ is Boltzmann's constant and $\Delta$ is the superconducting gap. Below $T = \SI{225}{\milli K}$, we are unable to accurately read out an exponential decay in $I$ and $Q$ due to the low QP density (see Appendix G). Starting $T = \SI{225}{\milli K}$, we observe the $t_\mathrm{trap}$ and $t_\mathrm{clear}$ decay as a function of $T$ as seen in \cref{fig:fig5} (e) reaching values of a few \SI{100}{\nano s}. 
The trends observed in \cref{fig:fig5} (e) as a function of $T$ are similar to that of \cref{fig:fig5} (d). 
Here, the suppression of $t_\mathrm{trap}$ and $t_\mathrm{clear}$ can also be attributed to the increasing recombination rate with increasing thermal quasiparticle density as well as to the increased phonon population at higher temperatures and phonon emission generated by QP recombination and QP inelastic scattering. As with the $V_\mathrm{inj}$ case, the recombination and excitation of trapped QPs through phonon absorption are  responsible for the decrease of $t_\mathrm{clear}$ with $T$. Another contributing factor is the suppression of the induced gap at higher temperatures, resulting in the Andreev traps getting shallower. By fitting the temperature dependence of $t_\mathrm{trap}$ and $t_\mathrm{clear}$, we find that both fit well to $1/x_\mathrm{eq}(T)$ if the the superconducting gap is reduced by a factor of 1.5 as seen in \cref{fig:fig5}(e). The inverse dependence of $t_\mathrm{trap}$ and $t_\mathrm{clear}$ on $T$ can be understood by considering a phenomenological model for the QP density as described in Appendix H; however, further investigation is required to understand the origin of such superconducting gap suppression. The observed trends in trapping and clearing times as a function of $V_\mathrm{inj}$ and $T$ suggest that our system is not well described by a simple Markovian model; rather, the QP dynamics observed in such planar junctions fabricated on hybrid superconductor-semiconductor structures involve interdependent mechanisms and warrants further theoretical understanding. Future work can explore to what degree these systems act Markovian or non-Markovian. 

\section{Conclusion}

In conclusion, we have studied QP-induced losses as well as QP trapping and clearing dynamics in high-transparency Al-InAs planar Josephson junctions.
By examining the resonance lineshape as a function of local QP injection, temperature, and photon number, we observe lineshape broadening and frequency shifts consistent with quasiparticle trapping. Using a model for the excitation of trapped QPs find a factor of 15 increase in the QP trapping rate as we flux bias the SQUID from 0.0 to 0.5 flux, corresponding to trapping in the deepened ABSs. By performing time-domain measurements, we observe that the trapping and clearing rates occur on a time scale that varies from a few \SI{}{\micro s} to hundreds of \SI{}{\nano s} depending on the trap depth and quasiparticle density. The results indicate that the relaxation of quasiparticles relies on electron-phonon interactions. On the other hand, the excitation mechanisms of QP from the traps are observed to have contributions from microwave photons and phonon absorption. The trapping and clearing times depending on the QP density suggest the presence of phonon-mediated QP-QP interactions in the system. The results highlight the dynamics of QP poisoning and considerations that need to be taken regarding microwave loss and operation time scales when developing qubits, specifically topological qubits, on hybrid superconductor-semiconductor systems.

\section{Acknowledgements}

We thank Max Hays and Enrico Rossi for fruitful discussions. The N.Y.U. team acknowledge support from the Army Research Office agreement W911NF2110303 and W911NF2210048. The N.Y.U. team also acknowledge support from MURI ONR award no. N00014-22-1-2764 P00001 and from the National Science Foundation agreement 2340-206-2014878 A01. W.M.S. acknowledges funding from the ARO/LPS QuaCR Graduate Fellowship. The authors acknowledge MIT Lincoln Laboratory and IARPA for providing the TWPA used in this work. We thank S{\o}ren Anderson for providing the schematics shown in \cref{fig:qcage}.

\section{Appendix A: Tight binding calculations}

We conduct tight-binding simulations using the Kwant python package \cite{groth_kwant_2014} along the lines of the work presented in Ref. \citenum{moehle_controlling_2022}. The simulation considers a 2D system where a semiconductor region separates two superconducting leads. The system is described by the Hamiltonian: 

\begin{equation}
\begin{bmatrix}
H & \Delta(x)\\
\Delta(x)^* & -H
\end{bmatrix},~
 H = -\frac{\hbar^2}{2m^*}\nabla^2-\mu   
\end{equation} 
where $m^*$ is the effective electron mass,  and $\mu$ is the chemical potential. The superconducting pairing potential $\Delta(x)$ varies spatially as:

\begin{equation}
 \Delta(x)=
    \begin{cases}
        \Delta_0 & \text{if } x<-l/2\\
        0 &  -l/2 <x<l/2 \\
        \Delta_0 e^{i\phi} & \text{if } x>l/2
    \end{cases}
\end{equation}

where $\Delta_0$ is the superconducting gap, $l$ is the junction length and $\phi$ is the phase difference across the two superconducting leads. We discretize the Hamiltonian above on a square lattice with lattice constant $a = \SI{5}{\nano m}$ and simulate a system with $w = \SI{1}{\micro m}$, normal region length $l = \SI{100}{\nano m}$, geometrical length of the superconductor $l_{sc} = \SI{2}{\micro m}$ (which is larger than the coherence length of the proximitized 2DEG $\xi = \SI{1373}{\nano m}$ calculated using $\xi = \hbar \sqrt{2\mu/m^*}/\Delta$ using the following parameters), superconducting gap $\Delta_0 = \SI{220}{\micro eV}$, carrier density $n = 4 \times 10^{11}\SI{}{\centi m^{-2}}$, effective electron mass $m^* = 0.04m_{e}$ and $m_{e}$ is the electron mass. 
To calculate the Andreev spectrum, we diagonalize the Hamiltonian and plot the energy as a function of $\phi$. 

\section{Appendix B: Material growth and device fabrication}

\begin{figure}[htp]
    \centering
    \includegraphics[width=0.45\textwidth]{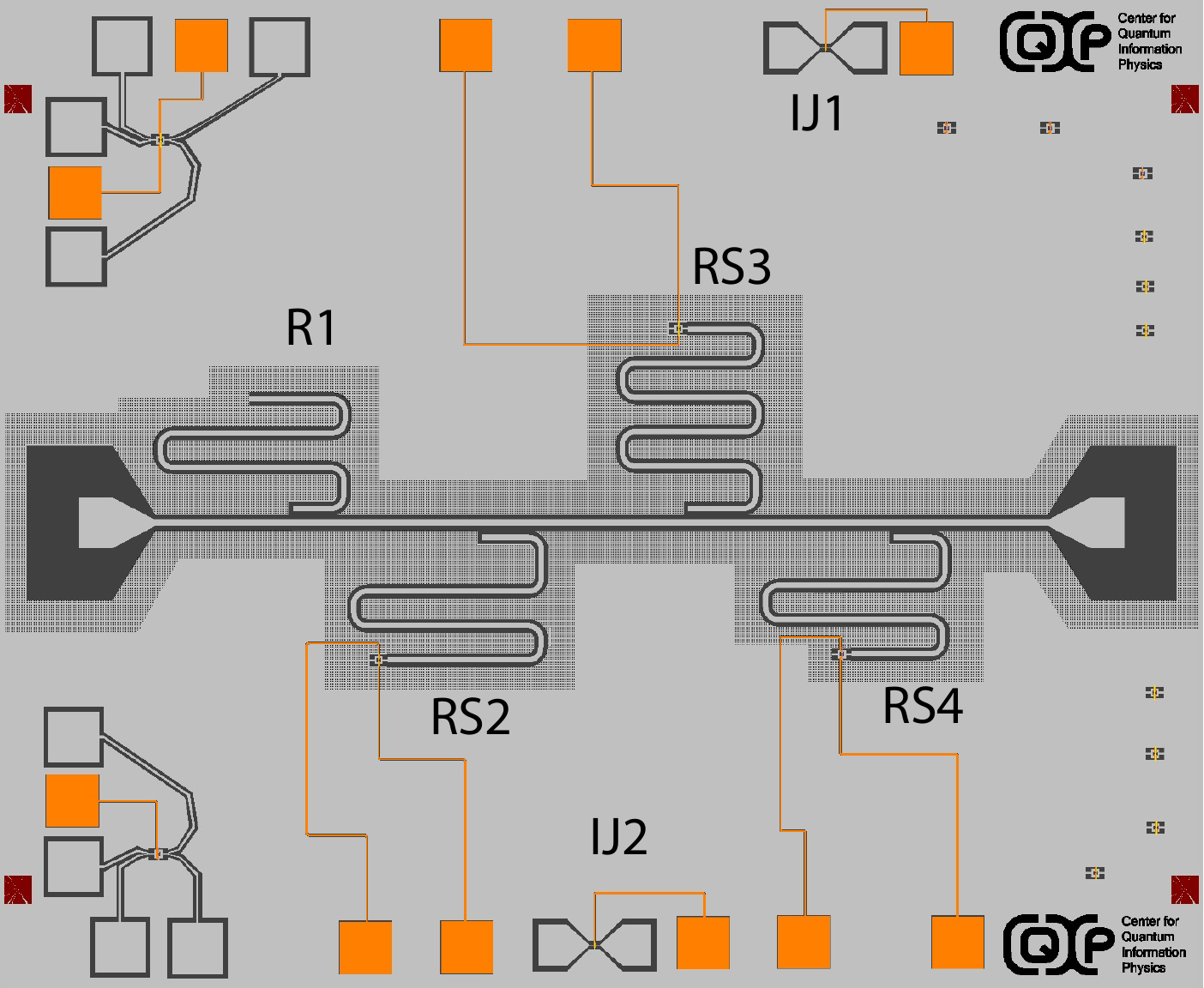}
    \caption{Schematic of the chip design. The chip has three $\lambda/4$ CPWs shunted to ground through a SQUID loop (RS2, RS3, and RS4) and one bare resonator (R1). The design also includes two injector junctions (IJ1 and IJ2). In this work, we focus on RS3, R1, and IJ1.} 
    \label{fig:design}
\end{figure}

\begin{figure}[htp]
    \centering
    \includegraphics[width=0.45\textwidth]{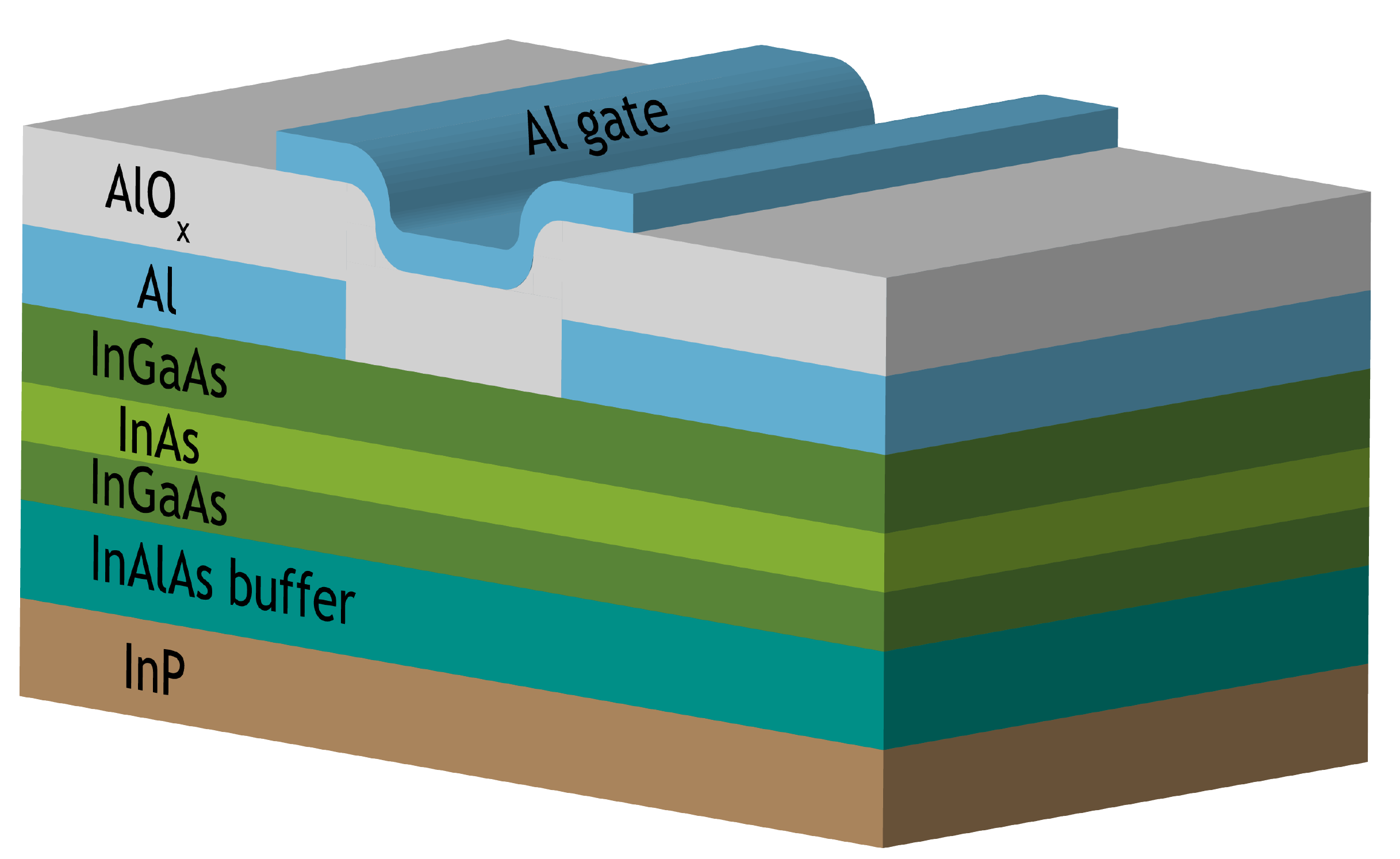}
    \caption{Schematic of the material heterostructure with a junction, made of Al superconducting contacts and an InAs surface quantum well. A layer of AlO$_\text{x}$ is deposited as a gate dielectric followed by a patterned Al gate.} 
    \label{fig:hetero}
\end{figure}

The CPW-SQUID devices are fabricated on an InAs near-surface quantum well grown by molecular beam epitaxy on a \SI{500}{\micro m} thick InP substrate. After thermal oxide desorption, an In$_{x}$Al$_{1-x}$As graded buffer layer is grown to reduce strain on the InAs active region, where the composition $x$ is graded from 0.52 to 0.81. The InAs 2DEG is confined to In$_{0.81}$Ga$_{0.19}$As top and bottom barriers. Finally, a \SI{30}{\nano m} film of Al is deposited \textit{in-situ}. The procedure for the growth of the III-V heterostructure is outlined in Refs. \citenum{Shabani2016, Kaushini2018, strickland2022}. 

The design was constructed using Qiskit Metal \cite{Qiskit_Metal} and rendered in Ansys's high frequency simulation software (HFSS) \cite{ansys} to simulate the expected resonant frequency, external quality factors, and electromagnetic field distribution. The chip design consists of four hanger CPW resonators coupled to a transmission line as seen in \cref{fig:design}. The external quality factor is designed to be $Q_\mathrm{ext}\sim 1500$. Three $\lambda/4$ CPWs are shunted to ground through a SQUID loop with two geometrically symmetric Josephson junctions. These devices are called CPW-SQUIDs (RS2, RS3, RS4). One bare CPW (R1), which does not have a SQUID loop and is shorted to ground directly, is used as a reference. We also add two two-terminal Josephson junctions as ``injector'' junctions (IJ1, IJ2) to inject QPs into the circuit. The injector junction is equipped with a gate, as shown in \cref{fig:fig1}(f), that is kept grounded during the measurements. In this work, we focus on RS3, R1, and IJ1.

The device is fabricated with electron beam lithography steps using spin-coated PMMA resist. First, we define the microwave circuit and chemically etch the Al using Transene Al etchant type-D and the III-V layers using an III-V etchant consisting of phosphoric acid (H$_3$PO$_4$, 85\%), hydrogen peroxide (H$_2$O$_2$, 30\%) and deionized water in a volumetric ratio of 1:1:40. The junctions in the CPW-SQUID are defined to be $\sim\SI{4}{\micro m}$ and the injector junction to be $\sim\SI{8}{\micro m}$. The planar junction is then defined by etching a thin \SI{100}{\nano m} strip of Al. Considering an electron mean free path of $\sim\SI{200}{\nano m}$ as measured by low-temperature Hall measurements, the junction should be in the short ballistic regime. Around the microwave circuit, we also etch $2\times\SI{2}{\micro m}$ holes in the Al ground plane that are $\SI{10}{\micro m}$ apart to act as flux-pinning holes. We then blanket deposit a \SI{40}{\nano m} layer of AlO$_\text{x}$ as a gate dielectric by atomic layer depositions, followed by a sputtered gate electrode made of \SI{50}{\nano m} Al layer using liftoff. In all measurements, the gates are kept grounded. A schematic of the junction heterostructure after fabrication is shown in \cref{fig:hetero}.

\section{Appendix C: Measurement setup}

\begin{figure*}[htpb]
    \centering
    \includegraphics[width=0.85\textwidth]{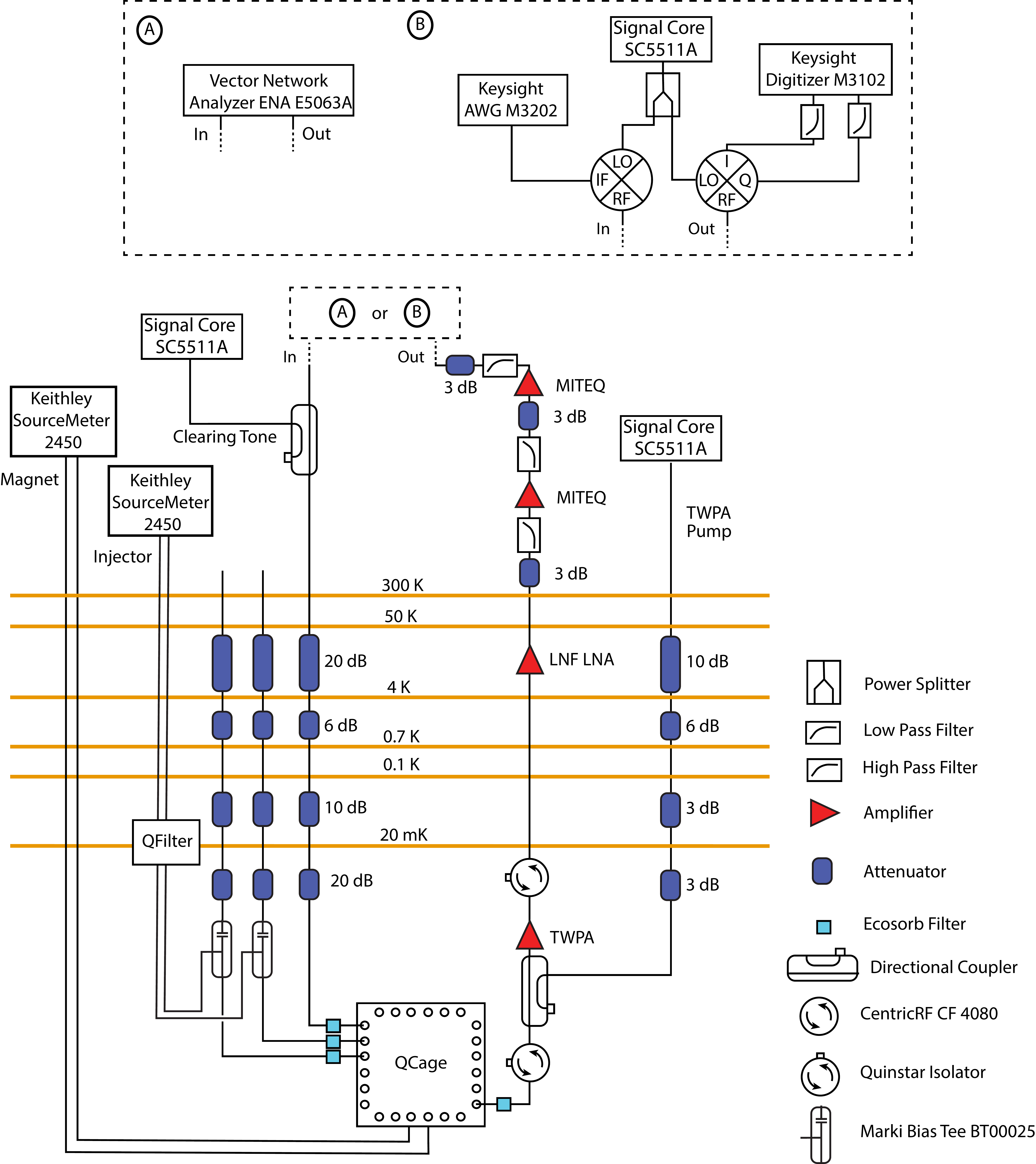}
    \caption{Schematic of the cryogenic and room temperature measurement setup.}
    \label{fig:fridge}
\end{figure*}

A schematic of the measurement setup is shown in \cref{fig:fridge}. Measurements are conducted in an Oxford Triton dilution refrigerator. The sample is embedded in a QCage, a microwave sample holder, and connected to the printed circuit board by aluminum wirebonds. Probe signals are sent from a vector network analyzer or a microwave signal generator and attenuated by -56 dB with attenuation at each plate as noted. The signal then passes through a 1-18 GHz bandpass filter made of a copper box filled with cured Eccosorb castable epoxy resin. The signal is sent through the sample, returned through another Eccosorb filter, and passed through an isolator with 20 dB isolation and 0.2 dB insertion loss. The signal is then amplified by a traveling wave parametric amplifier, passed through another isolator, and then amplified with a low noise amplifier mounted to the 4K plate and two room temperature amplifiers (MITEQ) outside the fridge.

We utilize the injector junction to increase the QP density of the system by voltage-biasing it above twice the superconducting gap. The two-terminal JJ is biased by grounding one terminal and applying a voltage bias to the other terminal. As mentioned in the text, all the DC lines in the fridge go through a QDevil QFilter, a low-pass filter that is mounted on the mixing chamber. The filters add a resistance of $\SI{1.7}{\kilo \Omega}$ to each DC line. Our reported applied voltage bias values to the injector junction $V_\mathrm{inj}$ is applied across the filter and the injector junction in series. Given that the normal resistance of the injector junction is unknown, it is difficult to determine the actual value of the voltage applied to or the power dissipated by the injector junction. However, we can make a simple approximation by noting that the data in \cref{fig:fig4} and the fact that no noticeable jumps in I and Q were observed when applying the clearing tone below $V_\mathrm{inj} = \SI{2}{\milli V}$, imply that considerable QP injection starts around $V_\mathrm{inj} = \SI{2}{\milli V}$. This would correspond to the actual voltage bias reaching the injector junction being around $2\Delta_{0} \sim \SI{440}{\micro eV}$ at $V_\mathrm{inj} = \SI{2}{\milli V}$. In that case, the normal resistance of the injector junction would be $\sim\SI{375}{\Omega}$, which is similar to what has been reported in other works on Al-InAs junctions with similar geometries \cite{dartiailh_missing_2021, elfeky_local_2021}.

\section{Appendix D: Flux calibration}

\begin{figure}[htp]
    \centering
    \includegraphics[width=0.5\textwidth]{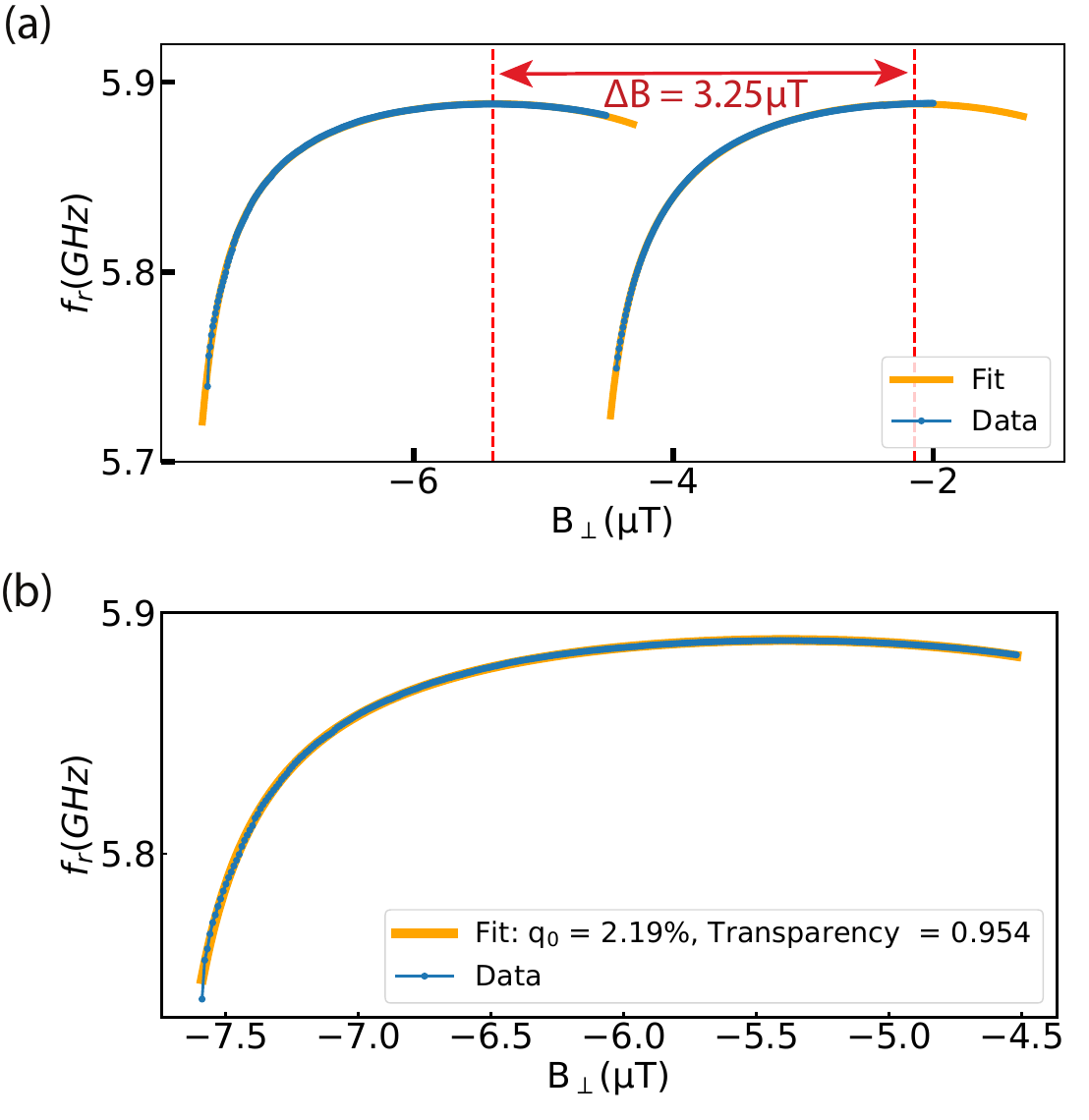}
    \caption{(a) Extracted resonant frequencies $f_{r}$ from \cref{fig:fig2}(a) as a function of the out-of-plane magnetic field $B_{\perp}$. The periodicity of the oscillations is found by finding the $B_{\perp}$ values at the maximum of each oscillation. (b) The extracted resonant frequencies and fit for the SQUID oscillation using \cref{eq:freq}. } 
    \label{fig:fieldfit}
\end{figure}

For the CPW-SQUID device, the SQUID is expected to show a periodic dependence as a function of $\Phi$, with periodicity $\Phi_{0}$, which should be reflected in the resonant frequency of the resonator. To extract the periodicity of the SQUID oscillations and calibrate the x-axis from the out-of-plane magnetic field $B_{\perp}$ to $\Phi/\Phi_{0}$ as shown in \cref{fig:fig2}(a), we first extract the resonant frequency of the data presented. The extracted resonant frequencies of the oscillations in \cref{fig:fig2}(a) are shown in \cref{fig:fieldfit}(a) as a function of $B_{\perp}$. We then find the $B_{\perp}$ point corresponding to the maximum frequency for each of the two oscillations and the difference in $B_{\perp}$ between these two points $\Delta B_{\perp} = \SI{3.25}{\micro T}$ is considered to be the periodicity of the SQUID oscillations. To calibrate the x-axis from $B_{\perp}$ to $\Phi/\Phi_{0}$ as shown in \cref{fig:fig2}(a), we subtract the value of $B_{\perp}$ for the maximum frequency of the left oscillation and divide the axis by $\Delta B_{\perp}$.

We fit the SQUID oscillations shown in \cref{fig:fieldfit}(a) to extract details of the SQUID device. Similar to the approaches taken in Refs. \citenum{LevensonFalk2013StaticAM,farmer2021}, the resonant frequency can be described by:

\begin{equation}
    \omega(\phi) = \omega(0)[1+q_{0}\frac{L_{J}(\phi)-L_{J}(0)}{L_{J}(0)}]^{-1/2}
    \label{eq:freq}
\end{equation}

with 

\begin{equation}
  I_\mathrm{ABS} (\phi) = \frac{e\Delta_{0}}{2\hbar}\frac{\tau \sin(\phi)}{\sqrt{1-\tau \sin(\phi/2)}}
\end{equation}

where $q_{0} = L_{J}(0)/(L_{0}+L_{J}(0))$ is the participation ratio and $\Delta_{0} = \SI{210}{\micro eV}$ is the superconducting gap . Fitting the left squid oscillation to \cref{eq:freq}, as presented in \cref{fig:fieldfit}(b), results in periodicity $\Delta B_{\perp} = \SI{3.4}{\micro T}$ with a participation ratio of $q_{0} = 2.19\%$ and transparency $\tau = 0.954$ as seen in \cref{fig:fieldfit}(b). This $\Delta B_{\perp}$ value is close to what was determined from \cref{fig:fieldfit}(a) by comparing the maximum frequency points of two oscillations.

We note that the SQUID studied in the CPW-SQUID device exhibits unusual flux tunability. Typically, SQUIDs with tunnel junctions exhibit a minimum frequency close to $\Phi/\Phi_{0} = 0.5$. Compared to the sinusoidal CPR exhibited by tunnel junctions, high-transparency superconductor/semiconductor junctions can exhibit a non-sinusoidal CPR. Such a non-sinusoidal CPR can result in the resonant frequency going beyond $\Phi/\Phi_{0} = 0.5$ as was discussed in Refs. \citenum{levenson-falk_nonlinear_2011,LevensonFalk2013StaticAM}. In \cref{fig:fig2}(a), the resonant frequency of the CPW-SQUID is seen to extend past $\Phi/\Phi_{0} = 0.5$, which we attribute to the non-sinusoidal CPR of our highly-transparent Al-InAs JJs.

\section{Appendix E: Non-equilibrium QP density}

As mentioned in the main text, without QP injection, i.e. $V_\mathrm{inj} = \SI{0}{\milli V}$, and at low temperature, we do not observe a QP clearing effect with the application of the clearing tone. We also confirm that with time-domain measurements, we are unable to see jumps between QP occupation states corresponding to QP poisoning events in real-time as was observed in Refs. \citenum{Hays2020, farmer2021}. Several factors may contribute to this, which include a low sensitivity to QP poisoning and a low QP density near the SQUID.

\subsection{Sensitivity to QP poisoning} 

For the CPW-SQUID, the junctions add a total inductance $L_{J}$/2 in series to the CPW, where $L_{J}$ is the inductance of each junction. The measured resonant frequency of the CPW-SQUID device is therefore determined by $L_{0}$, $L_{J}$, and the kinetic inductance of the thin Al film $L_{K}$. We use a bare resonator on the same chip to estimate $L_{K}$ of the thin film Al superconductor. For the bare resonator, we calculate the kinetic inductance fraction $\alpha_K = 1-({f_{r}}/f_{0})^2 = 7.27 \%$, where ${f_{r}}$ and ${f_{0}}$ are the measured and geometric resonant frequencies, respectively. This sizable kinetic inductance is expected for the thin epitaxial Al film. $L_{K}$ of the CPW-SQUID is then given by $L_{0}\alpha_K$ where $L_0$ is determined using an analytical expression for the inductance of a $\lambda/4$ CPW using conformal mapping techniques \cite{simons_CPW}. Knowing $L_{0}$, $L_{K}$ and $f_{r}$ allows for the determination of the $L_{J}$ in the CPW-SQUID device as described further in Ref. \citenum{strickland_superconducting_2023}.

At $\Phi/\Phi_0 = 0.5$, the CPW-SQUID device has a resonant frequency of $f_r = \SI{5.860}{\giga Hz}$ and a linewidth of \SI{8.455}{\mega Hz}. This corresponds to a Josephson inductance of $L_{J}(\Phi/\Phi_{0}=0.5) = \SI{0.190}{\nano H}$ and a critical current of $I_{c}(\Phi/\Phi_{0}=0.5) = \SI{1.73}{\micro A}$. Semiclassically, we can estimate the number of modes to be $N_{e} \sim w/\lambda_{F} \approx 280$, where $\lambda_{F}$ is the Fermi wavelength. Thus, the average current carried by a single mode is $I_{c}/N_{e} = \SI{6.19}{\nano A}$, and a corresponding single quasiparticle poisoning event would result in a frequency shift of $\sim \SI{0.28}{\mega Hz}$ on average. Practically, we expect deeper traps to have a higher chance of occupying QPs (assuming electron-phonon interactions, the rate of QP trapping can be taken to be $\propto \Delta_{A}^{3}$ \cite{kaplan_quasiparticle_1976, marchegiani_quasiparticles_2022,farmer2022} near $\Delta_{0}$ by solving the kinetic equation for a quasiparticle distribution) and the supercurrent to be carried mainly by high transparency modes as implied by \cref{fig:fig1}(b). For a mode in the ballistic limit carrying the maximal current ($\tau =1$) of $e\Delta/\hbar \approx\SI{51}{\nano A}$, a quasiparticle poisoning event results in a shift of $\sim\SI{2.46}{\mega Hz}$. Thus, due to the wide distribution of transparencies, we expect QP poisoning events to result in a frequency shift between these two values per trapped QP. Given our signal-to-noise ratio, it could be the case that shifts due to QP trapping are considerably less than the linewidth of the resonance. Consequently, without QP injection and at low temperatures, the relatively ``low'' number of trapped QPs would not result in a significant effect on the resonance shape when the clearing tone is applied or allow us to observe single QP trapping in real-time. However, with QP injection and at higher temperatures, the QP density is high enough that a sufficient amount of QPs are trapped such that the application of the clearing tone has a noticeable effect on the resonance shape. 

\subsection{Low nonequilibrium QP density} 

\begin{figure}[htp]
    \centering
    \includegraphics[width=0.5\textwidth]{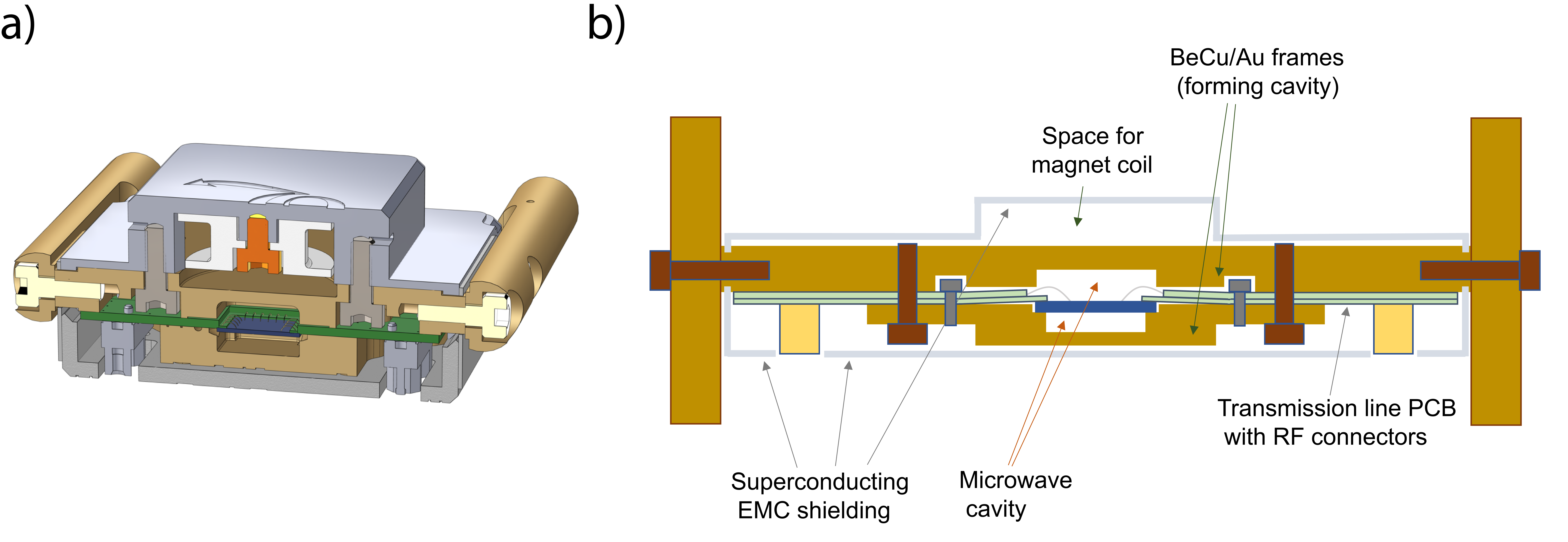}
    \caption{(a) 3D diagram of the Qcage sample holder. (b) Cross-sectional view of sample holder showing the BeCu/Au frames that create the inner cavity and the superconducting shielding.} 
    \label{fig:qcage}
\end{figure}

A low density of nonequilibrium QPs near the junction could also account for the inability to measure real-time trapping of QPs. There are a number of reasons why this may be the case. First we consider the microwave packaging. For the measurements presented in this work, we use a Qcage, a microwave sample holder from Qdevil. The Qcage provides EMC tight shielding by two layers of shielding: a BeCu/Au frame that forms a cavity and a sealed superconducting enclosure as seen in \cref{fig:qcage}. The BeCu/Au inner cavity parts are designed to have no line of sight into the cavity. The machined vent channels are all routed to prevent direct lines of sight. The only direct line of sight into the cavity around the chip is through the dielectric layers of the PCB stack and the coaxial cables. The chip, PCB, and cavity parts are sealed with an aluminum enclosure that superconducts at low T. The only line of sight through the aluminum casing is via the BeCu/Au side mounting rods and the coaxial connectors. We speculate that this EMC-tight shielding could contribute to a low nonequilibrium QP density. 

Another reason contributing to a low nonequilibrium QP density around the CPW-SQUID could be the flux pinning holes surrounding the microwave circuit. Since the flux pinning holes are holes made in the superconducting ground plane, the exposed area of the holes is the conducting top InGaAs layer. This normal metal layer can act as a QP trap. Further, vortices pinned in the flux-pinning holes can also trap QPs. These two effects can reduce the QP density around the CPW-SQUID. 

We also consider the superconducting gap profile of the structure, where the Al layer is sitting on a proximitized layer of an InAs 2DEG. The proximitized InAs has an induced superconducting gap lower than superconducting parent gap of Al. QPs can then relax into the proximitized InAs occupying the states close to the induced gap, which could play a role in the reduction of the QP density in the Al layer (and correspondingly the JJ) \cite{marchegiani_quasiparticles_2022, connolly_coexistence_2023}. Finally, we note that the piezoelectricity of the InP substrate could absorb far infrared radiation. Further investigation is required to determine with certainty the origin of the lack of QP effects we see without deliberate QP injection.

\section{Appendix F: Efficiency of clearing tone}

\begin{figure}[htp]
    \centering
    \includegraphics[width=0.5\textwidth]{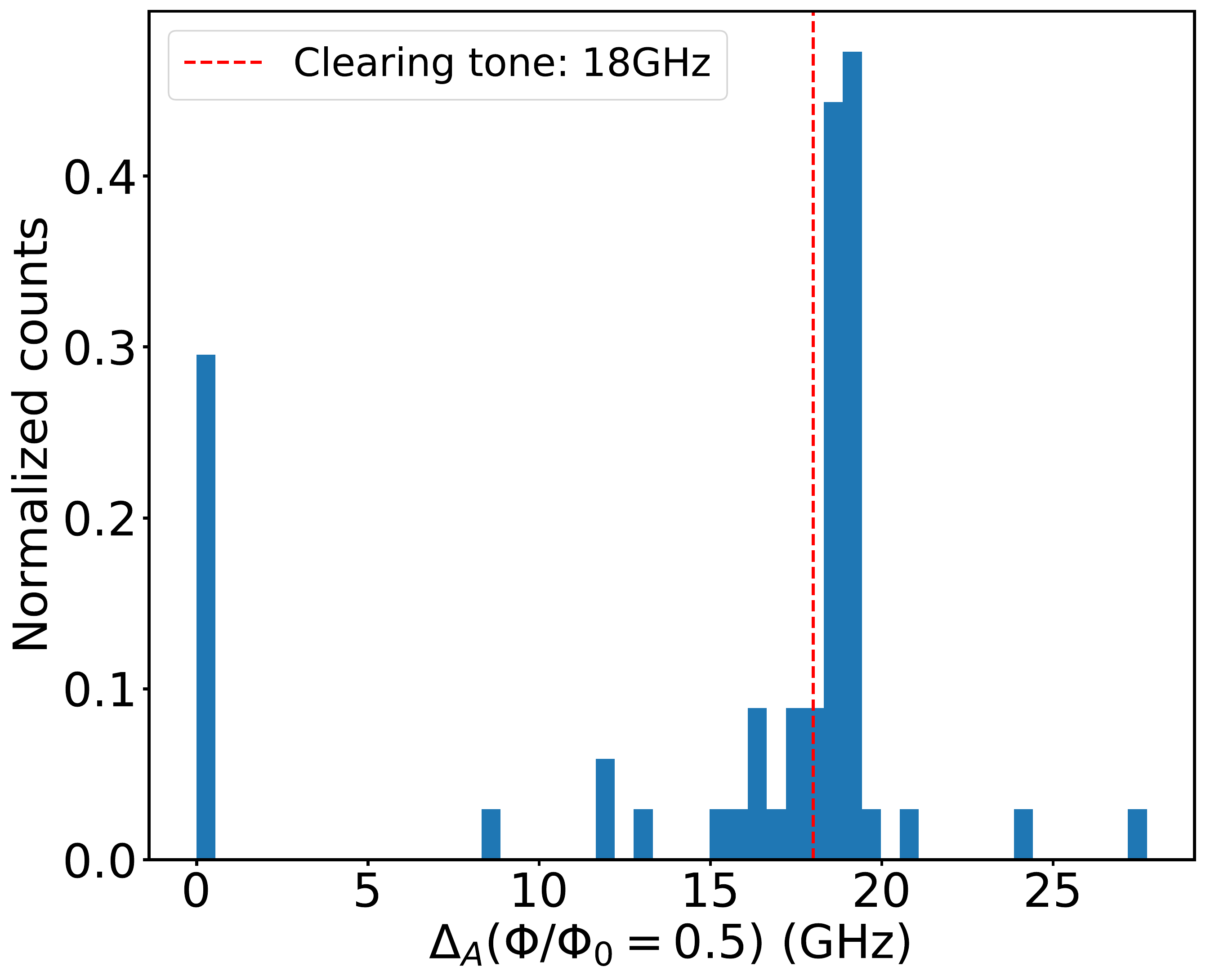}
    \caption{A histogram of the Andreev trap depth $\Delta_{A}$ at half-flux extracted from the ABS modes calculated in \cref{fig:fig1}(a). } 
    \label{fig:trapdepth}
\end{figure}

In this work, we utilize a high-frequency clearing tone to excite QP out of their traps. As we mention in the main text, the frequency chosen for the clearing tone is $f_\mathrm{clear} = \SI{18}{\giga Hz}$ since it is near the third harmonic of the $\lambda/4$ resonator where the admittance of the CPW is peaked. Ideally, the clearing tone should have an energy greater than the deepest Andreev trap, $h f_\mathrm{clear}>\Delta_{A} (\tau = 1)$, to clear QPs from all Andreev traps directly to the continuum. At half-flux, \cref{eq:gen_abs} results in a maximum $\Delta_{A}(\tau = 1)/h = \SI{14.87}{\giga Hz}$. However, \cref{eq:gen_abs} is a simple approximation for a 1D conduction channel. If we consider the ABS modes plotted in \cref{fig:fig1}(a) calculated using a 2D tight-binding model and we histogram the trap depth $\Delta_{A}$ at half-flux, presented in \cref{fig:trapdepth}, we see that a large number of the modes have a trap depth $\Delta_{A}/h\gtrsim\SI{18}{\giga Hz}$. A few outliers are also present that correspond to long junction modes, with a trap depth that is significantly larger than \SI{18}{\giga Hz}. The clearing of QP trapped in these modes with the applied \SI{18}{\giga Hz} will not be completely efficient and might involve mode-to-mode transitions rather than direct mode-to-continuum clearing of QPs.

Moreover, the results shown in \cref{fig:fig5} show that the clearing and trapping times are of similar magnitude. As discussed in the text, a steady state can be achieved where the clearing and trapping events balance each other, with the Andreev states constantly being partially occupied even in the presence of the clearing tone. This finite QP occupation could be partly responsible for the inefficiency in the clearing tone.

\section{Appendix G: Limitations in $t_\mathrm{trap}$ and $t_\mathrm{clear}$ extraction}

In \cref{fig:fig5}, we present time-domain measurements related to the trapping and clearing of QPs. As described in the text, $t_\mathrm{trap}$ and $t_\mathrm{clear}$ are extracted by fitting the clearing tone induced jumps and drops in I or Q to an exponential function \cite{levensonfalk2014}. At $V_\mathrm{inj} < \SI{2}{\milli V}$ and $T < \SI{225}{\milli K}$, given our sensitivity to QP poisoning events, the QP density in the system is low enough that there is not a significant amount of QPs occupying the Andreev traps. Therefore, the response in I and Q to the clearing tone is either not visible or very minimal, given our signal-to-noise ratio, which does not allow for a reliable fit to extract clearing or trapping times. Hence, we only presented data for $V_\mathrm{inj} \geq \SI{2}{\milli V}$ and $T \geq \SI{225}{\milli K}$ and above where the jumps and drops in I and Q are prominent enough that we can get a reliable fit to extract the trapping and clearing times.

\section{Appendix H: Phenomenological model for QP density}

In the following, we present a phenomenological model to describe trapping and clearing times' dependence on the quasiparticle density. We use the reduced QP density $x(t)$ in the superconducting region of the device in the vicinity of the Josephson junctions, $x = n_{QP}/n_{CP}$, where $n_{QP}$ is the density of QPs normalized to the density of Cooper pairs $n_{CP}$.
The equation for the reduced QP density has the form \cite{wang_measurement_2014, rothwarf_measurement_1967}:
\begin{equation}
\label{eq:rothwarf}
\dot x = - rx^2-sx -px +g.
\end{equation}
Here, $r$ and $s$ are the recombination and trapping rates for QPs in the superconducting region, $p$ is the trapping rate for the Josephson junction, and $g$ is the QP generation rate due to Cooper pair breaking by phonons or excitation of trapped QP by the photons in the resonator \cite{patel2017,grunhaupt2018}.

In our analysis, we assume that the reduced quasiparticle density $\bar x$ is stationary and according to \cref{eq:rothwarf}, depends on the QP generation rate $g$\begin{equation}
    \bar x = \frac{\sqrt{(s+p)^2+4gr}-s-p}{2r}.
\end{equation}
 However, when the injector is biased to inject QPs, 
we use $\bar x = \sqrt{g/r}$ for strong generation rate $gr\gg s^2$ and $\bar x = g/s$ for weak generation $gr\ll s^2$. Here, we assume $s\gg p$, that is the junction is not the main mechanism of quasiparticle trapping. The generation rate was estimated in Ref. \citenum{patel2017} as
$g\propto \sqrt{(eV_\mathrm{inj}/\Delta)^2-1}$.  Assuming that the quasiparticle generation rate is not high such that $gr\lesssim s$,  for $V_\mathrm{inj} \gg \Delta/e$ we have
\begin{equation}
    \label{eq:barx_inj}
    \bar x = \frac{g}{s} \propto V_\mathrm{inj} .
\end{equation}
When no QPs are injected by the injector junction, the QP density is defined by the thermal distribution, $\bar x = x_{\rm eq}(T)$, given by \cref{eq:eq_thermal}.

We also write the equation for the number of trapped quasiparticles in the ABSs of the junction $n_A$ as \cite{grunhaupt2018}:
\begin{equation}
\label{eq:dotNA}
\dot n_A = - r_{\rm qp,A}x n_A- r_{A} n_A^2  - \gamma_{\rm exc} n_A + pS x + g_A,
\end{equation}
where $r_{\rm qp,A}$ are the recombination rates of one quasiparticle in the superconducting region and one trapped quasiparticle in the junction, $r_A$ is the recombination of trapped quasiparticles, $S$ is the area of the junction, $\gamma_{\rm exc}$ is the excitation rate of trapped quasiparticles by the clearing tone, and  $g_A$ is the generation rate of trapped quasiparticles from other sources.  We again treat the rates in \cref{eq:dotNA} as phenomenological parameters.

We first evaluate the steady state value of the trapped quasiparticles.  We take $\dot n_A=0$ in Eq.~\eqref{eq:dotNA} and we obtain
\begin{equation}
\label{eq:barNA}
\bar n_A =\frac{pS \bar x + g_A}{r_{\rm qp,A}\bar x +\gamma_{\rm exc}}.
\end{equation}
Here, we assume that the recombination term $r_A n_A^2$ of trapped quasiparticles provides a negligible contribution.  
The number of photons in the resonator defines the excitation rate $\gamma_{\rm exc}$, and consequently, the number of trapped quasiparticles $\bar n_A$.  As the clearing pulse is turned on and off, the number of quasiparticles will switch between two values 
given by \cref{eq:barNA} with and without $\gamma_{\rm exc}$.

Next, we linearize \cref{eq:dotNA} in small changes of the trapped quasiparticle number:
\begin{equation}
\label{eq:Gamma}
    \delta \dot n_A = -\Gamma \delta n_A, \quad 
    \Gamma =  r_{\rm qp,A} x + 2 r_{A} n_A + \gamma_{\rm exc} .
\end{equation}
We can disregard  $2r_{A} n_A$ representing recombination of trapped quasiparticles.  We find that the trapping, $t_{\rm trap}$,  and clearing, $t_{\rm clear}$,  times are determined by the reduced quasiparticle density as 
\begin{equation}
    t_{\rm trap} =  \frac{1}{r_{\rm qp,A}\bar x}, \quad
    t_{\rm clear} =   \frac{1}{
    r_{\rm qp,A}\bar x +\gamma_{\rm exc}}.
    \label{eq:t_scaling}
\end{equation}
Because of the additional term $\gamma_{\rm exc}$ representing excitation of trapped quasiparticles by the clearing tone, the clearing time is shorter than the trapping time, $t_{\rm clear}<t_{\rm trap}$.  
In the absence of the injection voltage, $V_{\rm inj}=0$, the quasiparticle density $\bar x=x_{\rm eq}(T)$ is given by \cref{eq:eq_thermal}. The data shown in \cref{fig:fig5}(e) is consistent with the the scaling in \cref{eq:t_scaling} as both the trapping and clearing times are seen to scale with $x_{\rm eq}(T)$ given by \cref{eq:eq_thermal} if $\Delta$ is suppressed by a factor of 1.5. The reason for such gap is unclear and requires further investigation. With QP injection, we use $\bar x \propto V_{\rm inj}$ to fit the curves for the trapping and clearing times (see \cref{eq:barx_inj}) and find a good agreement with the data as shown in \cref{fig:fig5}(d).


\bibliography{References_Shabani_Growth}

\end{document}